\documentclass{article}

\usepackage[preprint, nonatbib]{neurips_2022}

\usepackage[utf8]{inputenc} 
\usepackage[T1]{fontenc}    
\usepackage{hyperref}       
\usepackage{url}            
\usepackage{booktabs}       
\usepackage{amsfonts}       
\usepackage{nicefrac}       
\usepackage{microtype}      
\usepackage{xcolor}         
\usepackage{amsmath}
\usepackage{adjustbox}
\usepackage{rotating}
\usepackage{multirow}
\usepackage{array}
\usepackage{graphicx}
\usepackage{enumitem}
\usepackage{soul}
\usepackage{ulem}
\usepackage{svg}
\usepackage{authblk}
\usepackage{float}
\usepackage{subfig}

\newcommand{\PreserveBackslash}[1]{\let\temp=\\#1\let\\=\temp}
\newcolumntype{C}[1]{>{\PreserveBackslash\centering}p{#1}}
\newcolumntype{R}[1]{>{\PreserveBackslash\raggedleft}p{#1}}
\newcolumntype{L}[1]{>{\PreserveBackslash\raggedright}p{#1}}

\title{Learning to Untangle Genome Assembly \\with Graph Convolutional Networks}

\author[1,2]{\textbf{Lovro Vrček}}
\author[3]{\textbf{Xavier Bresson}}
\author[4]{\textbf{Thomas Laurent}}
\author[1,3]{\textbf{Martin Schmitz}}
\author[1,2]{\textbf{Mile Šikić}}

\affil[1]{Genome Institute of Singapore, A*STAR}
\affil[2]{Faculty of Electrical Engineering and Computing, University of Zagreb}
\affil[3]{National University of Singapore}
\affil[4]{Loyola Marymount University \vspace{3px} \authorcr \tt \{vrcekl, miles\}@gis.a-star.edu.sg}

\begin{document}

\maketitle

\begin{abstract}

A quest to determine the complete sequence of a human DNA from telomere to telomere started three decades ago and was finally completed in 2021. This accomplishment was a result of a tremendous effort of numerous experts who engineered various tools and performed laborious manual inspection to achieve the first gapless genome sequence. However, such method can  hardly be used as a general approach to assemble different genomes, especially when the assembly speed is critical given the large amount of data. In this work, we explore a different approach to the central part of the genome assembly task that consists of untangling a large assembly graph from which a genomic sequence needs to be reconstructed. Our main motivation is to reduce human-engineered heuristics and use deep learning to develop more generalizable reconstruction techniques. Precisely, we introduce a new learning framework to train a graph convolutional network to resolve assembly graphs by finding a correct path through them. The training is supervised with a dataset generated from the resolved CHM13 human sequence and tested on assembly graphs built using real human PacBio HiFi reads. Experimental results show that a model, trained on simulated graphs generated solely from a single chromosome, is able to remarkably resolve all other chromosomes. Moreover, the model outperforms hand-crafted heuristics from a state-of-the-art \textit{de novo} assembler on the same graphs. Reconstructed chromosomes with graph networks are more accurate on nucleotide level, report lower number of contigs, higher genome reconstructed fraction and NG50/NGA50 assessment metrics. Both the code and the dataset are made publicly available\footnote{ \url{https://github.com/lvrcek/GNNome-assembly}}.
\end{abstract}

\section{Introduction}
\label{section:introduction}

\textit{De novo} genome assembly is a problem in bioinformatics that focuses on reconstructing the original genomic sequence of a species or an individual from a biological sample of shorter overlapping fragments, called reads, without any prior knowledge about the original sequence. There is a variety of tools that tackle this problem, so called \textit{de novo} genome assemblers, but complete reconstruction of large genome in a fast and accurate manner remains unsolved. One of the main issues is the fragmentation of the assembly genomes---instead of reconstructing each chromosome of the genome as a single contiguous sequence, it is broken into many shorter fragments, called contigs. The fragmentation happens when \textit{de novo} assembly tools fail to resolve a complex genomic region, which is, in lack of better methods, cut out from the reconstruction.

Researchers have been on a quest to assemble the complete human genome since 1990 \cite{lander2001initial}. A recent accomplishment, however, shows that a complete, unfragmented reconstruction of human genome is possible. This was enabled by the latest generation of sequencing technologies which produce longer and more accurate reads compared with those from previous generations, but also by a tremendous effort of numerous researchers who used different \textit{de novo} assembly tools to manually inspect remaining large repetitive regions and avoid fragmentation of the assemblies. To achieve this goal, first they used HiFi reads developed by Pacbio to reconstruct most of the genome, after which complex regions were resolved with ultra-long Nanopore reads.

Motivated by their effort, in this work we propose an automated method for reducing the fragmentation in \textit{de novo} genome assembly that relies neither on tedious human inspection nor inaccurate heuristics, but rather on graph neural networks. For this, we use HiFi reads only and focus on the central phase of \textit{de novo} genome assembly, in which a large, complex assembly graph has to be untangled into contigs. In our work we propose a new learning framework following the Overlap-Layout-Consensus assembly paradigm. Scientific contributions of this work are:
\begin{itemize}[leftmargin=*]
    \item \textbf{Assembly quality:} Starting from the same assembly graph, we reconstruct the chromosomes in fewer, longer fragments than heuristics commonly used in \textit{de novo} genome assembly.
    \item \textbf{Research direction:} To the best of our knowledge, this is the first ever work that utilizes deep learning to resolve the Layout phase of \textit{de novo} genome assembly. Thus, we open a new avenue to bring improvements over the current methods.
    \item \textbf{Dataset and framework:} We manually curate and annotate a dataset of human chromosomes used in the human genome reconstruction, which allows for training on real data and easier evaluation of models. We make the dataset publicly available, together with the code of the entire framework we implemented, from simulating the synthetic data to inferring on real data. 
\end{itemize}

\section{Background}
\label{section:background}

Contemporary \textit{de novo} assembly methods are primarily based on third generation long reads technologies. The most popular long reads sequencing technologies are produced by Pacific Biosciences (Pacbio) and Oxford Nanopore Technologies (ONT). Sequenced reads mutually vary in length and accuracy, so not all tools work equally good with both---there is no one-size-fits-all assembly tool. Although ONT devices can produce much longer reads (up to 1 million base pairs), we focus only on the PacBio HiFi reads due to combination of their high accuracy (above 99.5\%) and length (15000 - 25000 base pairs).

\paragraph{OLC} The Overlap-Layout-Consensus (OLC) paradigm is a popular approach to \textit{de novo} assembly with long reads, used also in the recent telomere to telomere reconstruction of the human genome.
In the Overlap phase, the reads are overlaped onto each other in an all-versus-all manner to find relations between them. Using the obtained information, an assembly graph is built---a directed graph in which nodes represent reads and edges represent the suffix-prefix overlaps between the reads. In the Layout phase, the assembly graph is then simplified in order to find a path through it. This path corresponds to the reconstructed sequence of the original genome, also called the assembly sequence. Finally, in the Consensus phase, the assembly sequence is cleaned of per-base errors by mapping all the reads onto the assembly.

\paragraph{Layout} Theoretically, this phase is formulated as finding a Hamiltonian path over the assembly graph---visit every node in the graph exactly once. However, due to errors in basecalling sequenced fragments, sequencing artifacts, long repetitive genomic regions, and heuristic overlap methods, constructed graph either does not have a Hamiltonian path or consists of spurious nodes or edges which should be skipped. Therefore, instead of finding a path through the graph directly, modern assemblers rely on heuristics to simplify the entire graph into a chain. This is done by iteratively removing nodes and edges deemed unnecessary, such as transitive edges, dead-ends, and bubbles \cite{li2016minimap, vaser2021time}. Frequently, however, some parts of assembly graphs, which usually correspond to highly complex genome regions, cannot be simplified by the current heuristics. In situations where a unique solution does not exist, contemporary assemblers cut out parts of these complex regions which leads to fragmented genome reconstructions. To this day, the problem of fragmentation continues to plague all the existing \textit{de novo} assemblers. The only remaining solution is laborious manual assembly inspection and curation.

\paragraph{Related work} So far, deep learning has been applied to different problems relevant to genome assembly, such as Consensus \cite{medaka}, basecalling for ONT reads \cite{wick2019performance}, and error-correction of HiFi reads \cite{baid2021deepconsensus},
in each case producing state-of-the-art results. However, to the best of our knowledge, no work has been done on solving the Layout phase with deep learning. The closest one is \cite{vrvcek2020step}, a proof-of-concept work where authors focus only on simulating the deterministic simplification steps in Layout, instead of untangling the graphs directly. Due to the type of the problem, it is natural to tackle Layout phase with Graph Neural Networks (GNNs) \cite{scarselli2008graph}, which have recently found a wide application in a variety of biological problems, ranging from drug design \cite{stokes2020deep} and protein interactions \cite{gainza2020deciphering}, to predicting anticancer foods \cite{gonzalez2021predicting}.

Hamiltonian Cycle Problem can be reduced to the Traveling Salesman Problem (TSP) by completing the graph and adding a unit cost to newly added edges while keeping zero cost for the existing ones. However, we cannot just plug-in existing state-of-the-art models for the TSP problem such as the graph transformer \cite{bresson2021transformer}, due to several reasons. First, assembly graphs are usually far larger than those used in the TSP research, often reaching hundreds of thousands of nodes and millions of edges. Second, for assembly graphs it is not possible to use coordinates of nodes as node features, like in the 2D TSP graphs that most of the research focuses on \cite{papadimitriou1977euclidean}. Finally, in \textit{de novo} assembly setting, spurious nodes (reads) and edges (overlaps), created due to imperfect sequencing procedure and overlap algorithms should be avoided. Consequently, we needed to conceive a GNN model carefully tailored for finding a path in assembly graphs. Nevertheless, we draw an inspiration from the work done on applying learning-based methods to TSP and follow a similar framework \cite{joshi2019efficient}.

\section{Problem setup}

\begin{figure}
\centering
\includegraphics[width=1\textwidth]{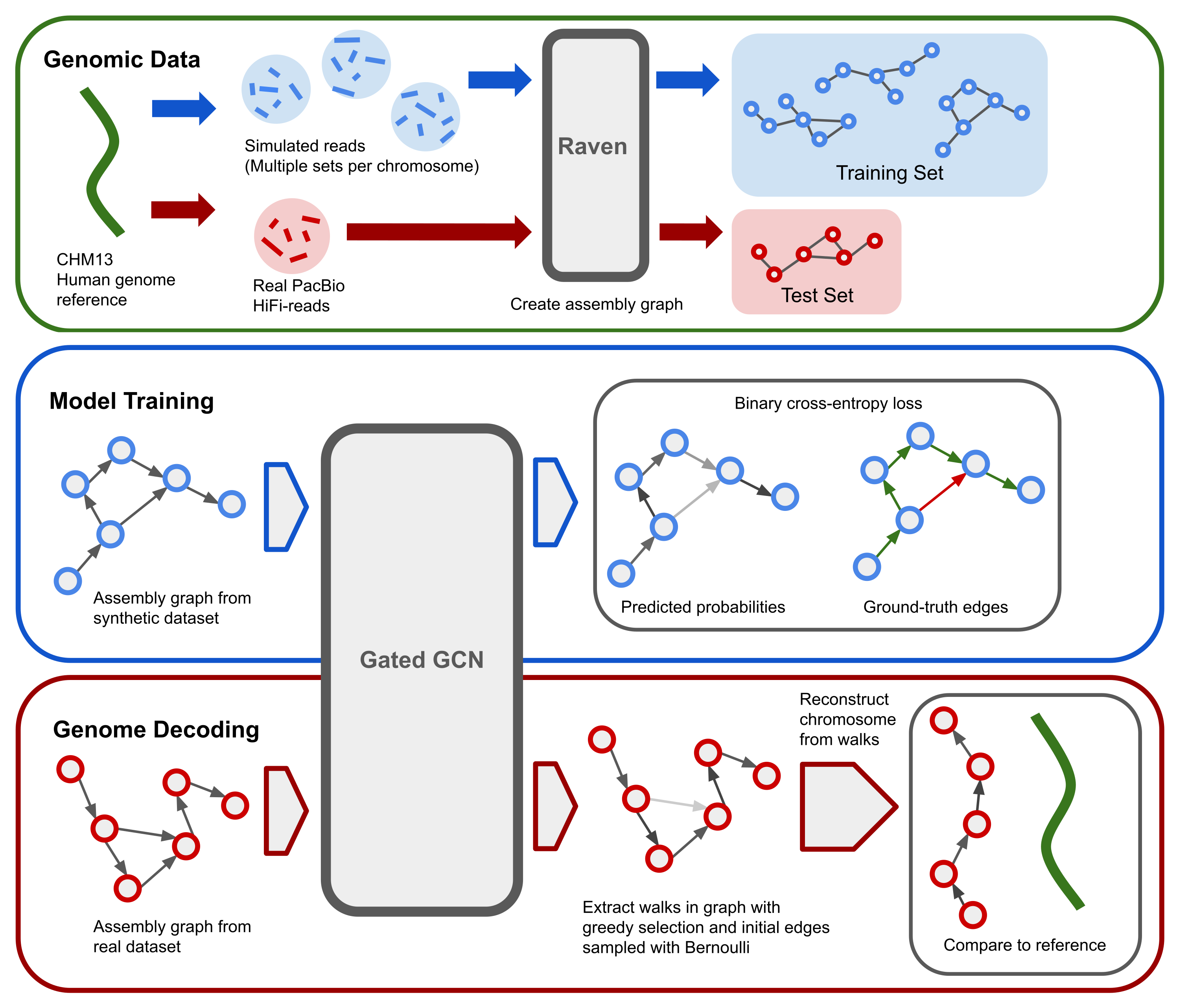}
\caption{Machine learning framework for end-to-end task of untangling \textit{de novo} assembly graphs and genome sequence reconstruction.}
\label{fig:framework}
\end{figure}

We propose the learning framework depicted in Figure \ref{fig:framework} to tackle the untangling of \textit{de novo} assembly graphs in an end-to-end manner.
The approach requires a previously completed genome, for which we use the recently reconstructed CHM13 human genome \cite{nurk2022complete}. It starts with simulating the reads from the CHM13 chromosomes from which assembly graphs are built using a \textit{de novo} genome assembler called Raven \cite{vaser2021time}.
Next step is training a model to predict which edges lead to the optimal assembly, running a search algorithm over those predictions, and translating the obtained paths into contigs. Once trained, we evaluate our model on the real PacBio HiFi reads used in the reconstruction of the CHM13 genome, and compare our approach to the heuristics Raven uses in the Layout phase. Since both methods have the same assembly graph at the input, this eliminates the effects different graph-construction techniques could have on the final assembly and enables us to compare only techniques for resolving Layout.

Raven is one of the state-of-the-art assemblers for long reads. Although there are assemblers tailored for Pacbio Hifi data, such as hifiasm \cite{cheng2021haplotype} and HiCanu \cite{nurk2020hicanu}, we chose Raven because, unlike many other assemblers, it doesn't specialize on a single type of reads, but performs well on both PacBio and ONT data. Even though in this work we focus on PacBio HiFi reads, we argue that the same approach can be applied to ONT data, which we plan to demonstrate in future research. Moreover, we argue that the same approach can be used not only on different types of long reads, but with any other OLC-based assembler.

\subsection{Dataset}
\label{subsection:dataset}

\subsubsection{Simulating synthetic reads}

We start with CHM13 reference genome \cite{nurk2022complete} which we first split into 23 chromosomes and sample them using a tool called seqrequester \cite{seqrequester}. Seqrequester samples the reference in such a manner that the length distribution of simulated reads faithfully resembles distribution of real HiFi reads. It also saves the positional information of each read, which facilitates distinguishing real from false overlaps in the graph. This process enables producing a large amount of data for supervised training, as well as constructing accurate labels for the edges. Although we start with only 23 chromosomes, by resampling we obtain a different set of reads each time, which result in graphs that are mutually sufficiently different to be considered as new elements. Although seqrequester produces errorless reads, due to high accuracy of HiFi data and error correction steps in the preprocesing, we deem that this would not result in a significant difference in the distribution of graphs created with real and simulated reads. The results prove our assumption.

\subsubsection{Preprocessing real reads}

We first error-correct reads with hifiasm \cite{cheng2021haplotype}, thus reducing amount of mismatches, insertions, and deletions in the reads. Then, we annotate the reads with their positional information by mapping them onto the reference with minimap2 \cite{li2018minimap2} and manually inspecting complex regions. Although annotations are not mandatory to evaluate the quality of the assembly, they serve us to compare how well will the model trained on synthetic graphs be able to predict edge-labels on the real-world graphs. This, additionally, gives a clue whether generalizing from synthetic to real-world data is possible. Annotating the real reads also facilitates using them for training, which could bring increase in performance. We deem this dataset will be important for future development and testing of similar learnable approaches to untangling assembly graphs, and thus make it available together with the code and present it in more details in Supplementary materials, Section \ref{supp:dataset}.

\subsubsection{Constructing graphs and labeling edges}

Once the reads are prepared, the assembly graphs are constructed using Raven's Overlap phase \cite{vaser2021time}. We output the generated graph prior to any simplification steps, in order to avoid errors that can occur during these steps. Due to heuristic nature of alignment algorithms used for overlapping the reads and complex genomic regions, there will exist edges in the graph which, if traversed, lead to suboptimal genome reconstruction. In order to train the model to avoid such edges, we assign them a label of 0, while all the edges which lead to optimal reconstruction are labeled as 1. We obtain these labels by running a DFS-like algorithm which finds the optimal path while taking into account the positional information saved during reads preparation. Due to transitive edges being present in the assembly graph, a large majority of edges will be labeled positively, thus making the dataset highly imbalanced towards the positive label. Eventually, this leaves us with a fully prepared dataset, on which a model can be trained for the task of binary edge classification. All the tools used in the described framework are publicly available and free to use---links to CHM13 dataset \cite{nurk2022complete} and the codes of Raven \cite{vaser2021time}, hifiasm \cite{cheng2021haplotype}, and minimap2 \cite{li2018minimap2} are available in their respective papers, while seqrequester is available at \url{https://github.com/marbl/seqrequester}.

\subsection{Model}
\label{subsec:model}


We train a Graph Neural Network called GatedGCN model \cite{bresson2017residual} that computes a $d$-dimensional representation of nodes and edges in the assembly graph, from which an MLP classifier outputs a probability that a given edge can lead to the optimal reconstruction. We compare these predicted probabilities to the correct edge labels obtained during the graph construction to calculate the binary cross entropy loss and minimize it with stochastic gradient descent. During inference, we do greedy search over the probabilities to find paths in the graphs, which are subsequently converted to contigs representing the reconstructed genome. The motivation for using GatedGCN comes from works previously done on the path-searching TSP problem \cite{joshi2019efficient,joshi2019learning} and a GNN benchmark where it outperformed other models on several tasks \cite{dwivedi2020benchmarking}.

\subsubsection{Input Features}
The input edge features $z_{ij} \in \mathbb{R}^{d_e}$ are composed of the length and quality of an overlap between two reads represented by two nodes of the assembly graph ($d_e=2$). They are normalized with a standard z-scoring (i.e. zero mean and unit standard deviation) over all edges.

We do not have any available genomic node features. In the absence of features, nodes are anonymous and GNNs are known to perform either poorly or completely fail to classify isomorphic graphs or detect basic patterns like cycles or cliques \cite{murphy2019relational, Loukas2020What, chen2020can}. The main reason of this failure is the lack of canonical positional information in arbitrary graphs. As such, recent works have proposed to augment node features with graph positional encoding such as \cite{dwivedi2021generalization, ying2021transformers, lim2022sign, dwivedi2022graph}. In the case of the directed assembly graph, we proposed to use the in-degree, out-degree and $k$-step diffused PageRank vector \cite{ilprints422} as node positional encoding. Note that these node positions are invariant to indexing permutation, which is essential for generalization. The $k$-step PageRank vector is generated with the following iterative scheme \begin{equation}
    p^{k+1} = \alpha (D^{-1}A)^T p^k + (1-\alpha)\frac{1_n}{n}\in\mathbb{R}^{n},\quad p^{k=0}=\frac{1_n}{n}\in\mathbb{R}^{n},
\end{equation}
where $A\in\mathbb{R}^{n\times n}$ is the adjacency matrix of the directed assembly graph, $n$ is the number of nodes, $D\in\mathbb{R}^{n\times n}$ is the diagonal matrix of the out-degree vector $A1_n\in\mathbb{R}^{n}$, $1_n\in\mathbb{R}^{n}$ is the $n$-dim vector of ones, and $\alpha=0.95$ is the random walker constant. To summarize, the input node features are $x_i=d^{\textrm{in}}_i \mathbin\Vert d^\textrm{out}_i \mathbin\Vert p_i^1 \mathbin\Vert \cdots \mathbin\Vert p_i^K \in \mathbb{R}^{d_v}$, where $\cdot\mathbin\Vert\cdot$ is the concatenation operator , $d_v=2+k$ with 2 for the degrees and $k$ for the dimentionality of the PageRank vector.

\subsubsection{Input layer}
An initial layer of the network transforms node features $x_i \in \mathbb{R}^{d_v}$ and edge features $z_{ij} \in \mathbb{R}^{d_e}$ into the $d$-dimensional node and edge representations.

This is done using two fully connected layers:
\begin{eqnarray}
    h_i^0 &=& W_{1,2}\mathrm{ReLU}(W_{1,1} x_i + b_{1,1}) + b_{1,2} \in\mathbb{R}^{d},\\
    e_{ij}^0 &=& W_{2,2}\mathrm{ReLU}(W_{2,1} z_{ij} + b_{2,1}) + b_{2,2} \in\mathbb{R}^{d},
\end{eqnarray}
where $h_i^0$ is the initial node representation of the node $i$, $e_{ij}^0$ is the initial representation of the edge $i \rightarrow j$, while all the $W$ and $b$ are learnable parameters.


\subsubsection{GatedGCN}
The main part of the network consists of multiple GatedGCN layers \cite{bresson2017residual}. In addition to the original GatedGCN, we include an edge feature representation $e_{ij}\in \mathbb{R}^d$, and use a dense attention map $\eta_{ij}\in \mathbb{R}^d$ for the edge gates, as proposed in \cite{bresson2019two, joshi2019efficient}. The original implementations, as most of the off-the-shelf GNN layers, are meant to be used on undirected graphs where messages are passed in all directions. However, in case of directed graphs this means that half of the messages are lost. For assembly graphs, which have an inherent directional information---from the beginning to the end of the genome---it is crucial to address this lack of expressivity.

Let $i$ and $p\rightarrow q$ be respectively the node and the directed edges (we will use $pq$ to denote $p\rightarrow q$ for simplicity) whose representations we want to update, and let their representations be $h_i^l$ and $e_{pq}^{l}$ at layer $l$. Also, let all the predecessors of node $i$ be denoted with $j$ and all its successors with $k$. Then, the node and edge representations at layer $l+1$ will be computed as:
\begin{eqnarray}
    h_i^{l+1} &=& h_i^{l} + \mathrm{ReLU} \bigg( \mathrm{BN} \bigg( A_1^l h_i^l + 
    \sum_{j \rightarrow i} \eta_{ji}^{f, l+1} \odot A_2^l h_j^l +
    \sum_{i \rightarrow k} \eta_{ik}^{b, l+1} \odot A_3^l h_k^l  \bigg) \bigg) \in\mathbb{R}^{d},\\
    e_{pq}^{l+1} &=& e_{pq}^{l} + \mathrm{ReLU} \Big( \mathrm{BN} \Big( B_1^l e_{pq}^{l} + B_2^l h_p^l + B_3^l h_q^l  \Big) \Big) \in\mathbb{R}^{d},
\end{eqnarray}
where all $A, B \in \mathbb{R}^{d \times d}$ are learnable parameters, ReLU stands for rectified linear unit, BN for batch normalization, and $\odot$ for Hadamard product. The edge gates are defined as:
\begin{equation}
    \eta_{ji}^{f, l} = \frac{\sigma\left( e_{ji}^{l} \right)}{\sum_{j' \rightarrow i} \sigma\left( e_{j'i}^{l} \right) + \epsilon} \in[0,1]^{d},\hspace{3em}
    \eta_{ik}^{b, l} = \frac{\sigma\left( e_{ik}^{l} \right)}{\sum_{i \rightarrow k'} \sigma\left( e_{ik'}^{l} \right) + \epsilon} \in[0,1]^{d}
\end{equation}
where $\sigma$ is the sigmoid function, and $\epsilon$ is a small value in order to avoid division by zero. Observe that we distinguish between the messages $\eta_{ji}^{f, l}$ passed along the edges, and in the reverse direction of the edges $\eta_{ik}^{b, l}$.

\subsubsection{MLP classifier} 


We use a multi-layer perceptron (MLP) on the node and edge representations produced by $L$ GatedGCN layers to classify the edges. For each directed edge $i \rightarrow k$, a probability $p_{ik}$ is computed and trained such that it leads to the optimal assembly, i.e. $p_{ik}=1$ if the edge is in the solution, or 0 otherwise. The probability is given by passing the concatenation of the node representation of nodes $i$ and $k$, as well as the edge representation of directed edge $i \rightarrow k$:
\begin{equation}
    p_{ik} = \sigma\big(\mathrm{MLP}\big(h_i^L \mathbin\Vert h_k^L \mathbin\Vert e_{ik}^{L}\big)\big)\in [0,1],
\end{equation}
where $\sigma$ is the sigmoid function and $L$ denotes the last GatedGCN layer.

\subsection{Sequence Decoding}
\label{subsection:decoding}

We decode the paths on the graph to assemble the reads with a greedy search algorithm. In the case where all the edge predictions were correct and the graph topology was noiseless (i.e. graph composed of a single connected component with neither dead-ends nor cycles) then extracting paths from the graph would become trivial by starting at any positively predicted edge and greedily choosing a sequence of edges with the highest probability in both forward and backward graph directions. However, neither of these conditions are met in practice. Hence, instead of performing a single greedy search, we first sample $B$ starting edges with Bernoulli, unroll $B$ greedy searches, and finally get the set of paths $\{w_1,\dots w_B\}$. Then, we compute the lengths of the sequences corresponding to the extracted paths, and choose the path with the longest sequence length. Subsequently, the selected path is translated into a contig of our reconstructed genome by concatenating the overlapping reads in the path. The nodes in the chosen path are masked out from the graph to avoid visiting them twice. The decoding process continues iteratively until the length of extracted path is below a certain threshold.

\section{Numerical Experiments}
\label{section:experiments}

\subsection{Model Training}
\label{subsec:training}

We first simulate 18 read-datasets from chromosome 19, from which assembly graphs are constructed. Out of those 18, we use 15 as the training set and the remaining as the validation set. There is nothing particularly special about chromosome 19. We chose it because it is one of the smaller chromosomes, while not being acrocentric. Acrocentric are chromosomes 13, 14, 15, 21, and 22, and are known to have more repetitive regions and be more difficult to assemble. This setup facilitates training at first, but it soon produced admirable results. As a sanity check, we additionally trained the model on five graphs from chromosomes 9, 19, and 22, each, but got only slightly better results, not statistically significant. Therefore, we report the results obtained with training and validating only on chromosome 19. The results achieved from training on a combination of chromosomes 9, 19, and 22 are available in the Supplementary materials, Section \ref{supp:mix-model}.


Since each set of reads was created anew by resampling the reference, they result in significantly different graphs, even if the reads are sampled from the same chromosome. This is confirmed by the fact that our model was able to generalize to different chromosomes with ease.

Due to their size, it was not possible to train the network on the entire graphs, so we partition the graphs using METIS clustering algorithm \cite{karypis1998fast}. We randomly choose a number of clusters from an interval between 400 and 600 each epoch. The obtained clusters are then grouped into mini-batches of size 50. We used Adam optimizer \cite{kingma2014adam}, with the initial learning learning rate of $10^{-3}$, and minimize the binary cross-entropy loss over each mini-batch. We decay the learning rate by a factor of 0.95 with patience of 2 epochs. Due to the dataset being highly imbalanced towards the positive class, we scale the learning rate for positive examples with an appropriate weight computed from the graphs in the training set. The entire training was done on a single Nvidia A100 GPU and took 53 minutes.

\paragraph{Hyperparameters} All the results reported in this paper are obtained with the network with hidden dimension $d=256$, number of GatedGCN layers $L=16$, and 3 layers in the MLP classifier, resulting in approximately 6.5 million parameters. They were obtained with a grid-search hyperparameter optimization over the mentioned hyperparameters. For the number of greedy path candidates, we choose $B=50$ because that is the lowest number of candidate paths that consistently produces optimal paths. We make the best performing pretrained model available together with the code.

\subsection{Inference}
\label{subsec:inference}

At inference, we do not use METIS to cluster the graphs in order to avoid cutting the edges from the graph. Instead, we feed the whole graph to the model, which in some cases requires us to run the model on a CPU, due to GPU memory limitations. However, the model is small enough so that a single forward pass of the network takes around 1 minute and 15 seconds for the largest graphs, and below 1 minute for the smaller ones on an AMD EPYC 7702 processor. This provides us with edge-probabilities which are used to guide the greedy search algorithm as described in Section \ref{subsection:decoding}, resulting in contigs which together form an assembly genome. Hence, we can evaluate our reconstruction using the same metrics that are commonly used in \textit{de novo} genome assembly. For this we use a tool called Quast \cite{gurevich2013quast}, and compare against several different measures:
\begin{itemize}[leftmargin=*]
\item \textbf{Number of contigs}: Gives an insight into how fragmented our reconstruction is (lower is better).
\item \textbf{Longest contigs}: The length of the longest contig (higher is better).
\item \textbf{Genome fraction}: Fraction of the genome which is reconstructed in our assembly (higher is better).
\item \textbf{NG50}: Length of the contig, which, coupled with longer contigs, covers 50\% of the reference genome (higher is better).
\item \textbf{NGA50}: Calculated the same way as NG50, but on top of alignments between contigs and the reference (higher is better).
\end{itemize}

\section{Results}
\label{section:results}


First we evaluate whether the model, trained only on graphs generated from simulated chromosome 19 reads, can generalize to other chromosomes constructed also from simulated reads. As an additional test, and to demonstrate the complexity of the problem, we run two na\"ive greedy searches, from the same $B$ starting edges as for decoding described in Subsection \ref{subsection:decoding}---one following the longest overlap and the other following the highest overlap similarity. Starting from the same edges removes the sampling randomness, and thus it is easy to see that the na\"ive approaches reconstruct significantly shorter contigs. Moreover, we show that the model can generalize to other chromosomes with a surprising ease, maintaining high performance on the entire dataset. When compared to Raven's Layout heuristics, GatedGCN outputs significantly fewer contigs for every chromosome, while having comparable reconstructed genome fraction. Moreover, GatedGCN creates contigs with either comparable or higher NG50 and NGA50 than the Raven's heuristics, in some cases improving them by more than 100\%. Full results of this experiment are available in the Supplementary materials Section \ref{supp:15xsynth}, where results of the na\"ive approaches are presented in Table \ref{table:baselines} and results of GatedGCN and Raven's heuristics are in Table \ref{table:15xsynthxcontig}. In the same Section, we compare the amount of mismatches and insertions/deletions (indels) per 100.000 base pairs of GatedGCN and Raven, and demonstrate that GatedGCN achieves fewer errors than Raven even in this comparison.

\begin{figure}
    
  \centering
  \includegraphics[width=\linewidth]{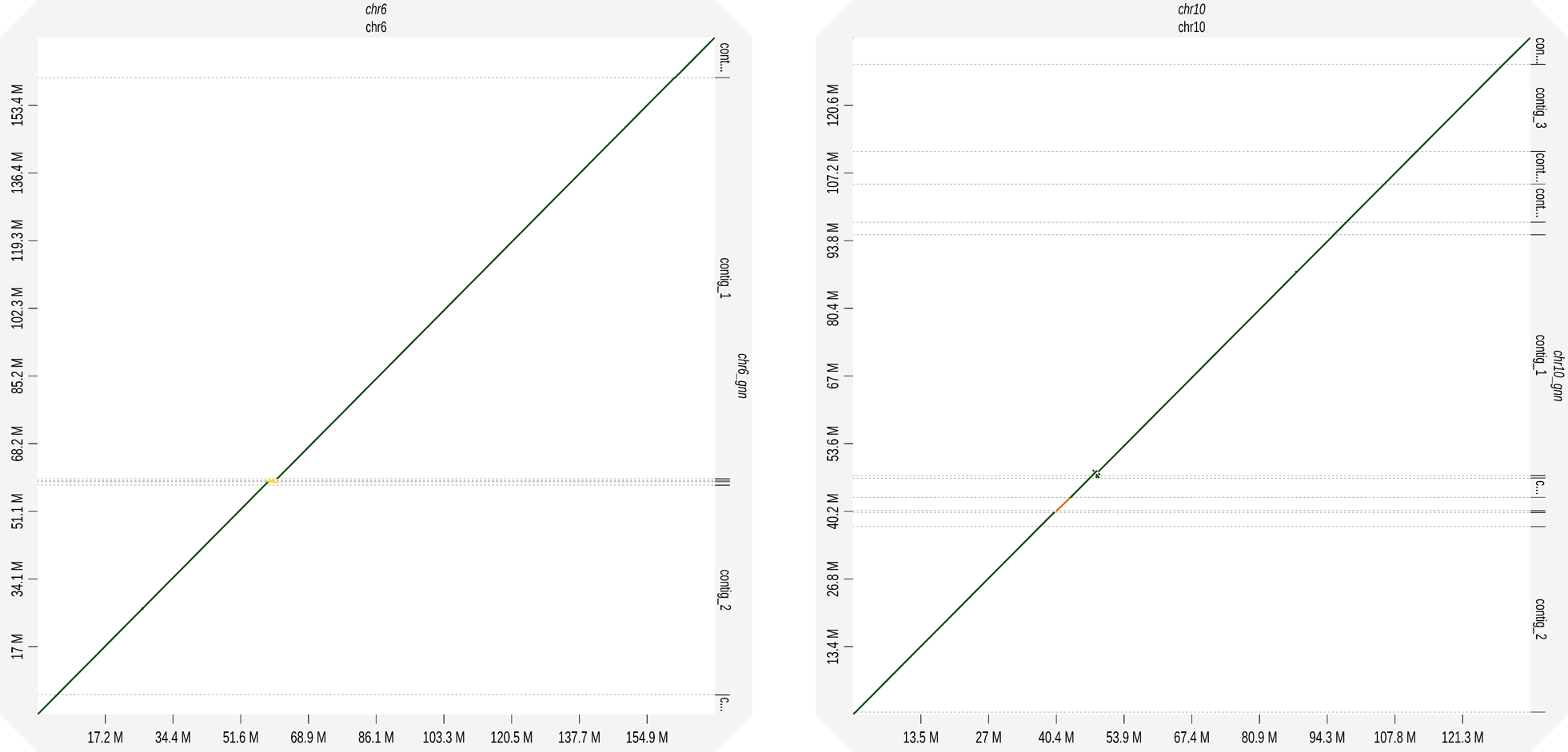}
  \caption{Our method's reconstruction of chromosomes 6 (left) and 10 (right). The plot shows the matching between the reference CHM13 chromosome on the $x$-axis and our reconstruction on the $y$-axis. Ideally, only a single dark green line would be visible on the diagonal. Horizontal dashed lines indicate fragmentation. Yellow and brown lines indicate low-quality mapping onto the reference, 0-25\% and 25-50\%, respectively, while light green and dark green lines represent 50-75\% and 75-100\% matching. Generated with DGenies \cite{cabanettes2018d}.}
  \label{fig:dgenies}
\end{figure}

After verifying that the network is able to generalize to synthetic graphs of other chromosomes, as well as outperform Raven's Layout heuristics, we evaluate whether the same prediction quality holds for assembly graphs generated from real human HiFi data. We assemble all the chromosomes, and compare our method with Raven's heuristics. These results are reported in Table \ref{table:exp3}.

Once again, our method outperforms Raven's heuristics on all of the reported tasks. It consistently produces fewer contigs, while the reconstructed genome fraction is in most cases comparable or higher---Raven's heuristics reconstruct more chromosome only in two cases, and the gain is less than 0.5\%. Longest contig, NG50, and NGA50 are also mostly comparable or higher, with just a few cases where Raven's heuristics outperform by a narrow margin. While NG50 is improved significantly for most of the chromosomes, most notable improvements in terms of NGA50 are seen in chromosomes 6 and 10, for which it is more than 100\% higher when compared to Raven's methods. Our reconstructions of these two chromosomes can be seen in Figure \ref{fig:dgenies}. When per-base errors are compared, presented method also mostly outperforms Raven's heuristics. Full results can be seen in Table \ref{table:15xrealxerror}.


\begin{table}
    \caption{Evaluation of our method (GatedGCN) and Raven's Layout heuristics on all the real chromosomes. The used metrics are number of contigs (Num ctg), longest contig (Longest), genome fraction (GF), NG50, and NGA50. With asterisk we denote chromosome 19 we used to simulate reads for the training dataset.}
    \label{table:exp3}
    \small
    \centering
    \begin{tabular}{cC{0.9cm}C{1cm}C{0.7cm}C{0.7cm}C{1cm}C{0.9cm}C{1cm}C{0.7cm}C{0.7cm}C{1cm}}
    \toprule
        & \multicolumn{5}{c}{\textbf{GatedGCN}} & \multicolumn{5}{c}{\textbf{Raven}} \\
        \cmidrule(r){2-6}  \cmidrule(r){7-11}
    \textbf{chr} & \textbf{Num ctg} & \textbf{Longest (Mbp)} & \textbf{GF (\%)} & \textbf{NG50 (Mbp)} & \textbf{NGA50 (Mbp)} & \textbf{Num ctg} &\textbf{Longest (Mbp)} & \textbf{GF (\%)} & \textbf{NG50 (Mbp)} & \textbf{NGA50 (Mbp)} \\
    \midrule

1 & \textbf{26} & \textbf{115.6} & \textbf{98.1} & \textbf{73.0} & \textbf{46.3} & 241  & 86.9 & 97.6 & 44.4 & 44.4\\
2 & \textbf{20} & 73.1 & \textbf{99.6} & \textbf{35.1} & \textbf{35.1} & 56  & 73.1 & 98.9 & 28.1 & 28.1\\
3 & \textbf{6} & \textbf{127.0} & \textbf{99.6} & \textbf{127.0} & 56.0 & 45  & 90.5 & 99.5 & 56.0 & 56.0\\
4 & \textbf{8} & \textbf{139.0} & 99.0 & \textbf{139.0} & 34.8 & 78  & 67.8 & 99.0 & 34.9 & \textbf{34.9}\\
5 & \textbf{8} & \textbf{123.6} & \textbf{99.1} & \textbf{123.6} & 103.5 & 47  & 103.5 & 99.0 & 103.5 & 103.5\\
6 & \textbf{7} & 101.0 & \textbf{98.9} & 101.0 & \textbf{52.8} & 20  & \textbf{110.3} & 98.7 & \textbf{110.3} & 25.9\\
7 & \textbf{17} & \textbf{58.6} & \textbf{98.1} & \textbf{42.6} & \textbf{25.7} & 69  & 29.3 & 98.0 & 25.1 & 17.5\\
8 & \textbf{12} & \textbf{68.8} & \textbf{98.6} & \textbf{33.9} & 28.5 & 33  & 31.6 & 98.4 & 28.5 & 28.5\\
9 & \textbf{17} & \textbf{67.1} & \textbf{95.0} & \textbf{31.9} & \textbf{16.1} & 139  & 38.9 & 90.2 & 19.7 & 15.8\\
10 & \textbf{13} & \textbf{47.7} & \textbf{99.3} & \textbf{36.7} & \textbf{36.7} & 43  & 36.7 & 99.2 & 17.2 & 17.2\\
11 & \textbf{7} & \textbf{65.4} & \textbf{99.9} & \textbf{35.3} & 23.2 & 31  & 35.3 & 99.7 & 32.6 & 23.2\\
12 & \textbf{11} & 57.2 & \textbf{99.9} & 31.0 & 31.0 & 33  & 57.2 & 99.8 & 31.0 & 31.0\\
13 & \textbf{13} & \textbf{73.0} & \textbf{96.1} & \textbf{73.0} & \textbf{30.1} & 116  & 47.5 & 95.9 & 25.5 & 25.5\\
14 & \textbf{9} & 82.6 & \textbf{97.8} & 82.6 & 82.6 & 32  & 82.6 & 97.2 & 82.6 & 82.6\\
15 & \textbf{19} & \textbf{47.1} & \textbf{93.6} & \textbf{13.4} & \textbf{10.0} & 157  & 29.0 & 93.5 & 9.0 & 8.5\\
16 & \textbf{28} & 16.0 & \textbf{91.6} & \textbf{8.7} & \textbf{5.9} & 164  & \textbf{16.4} & 90.8 & 5.9 & 5.7\\
17 & \textbf{11} & \textbf{29.9} & \textbf{96.4} & \textbf{15.7} & \textbf{10.2} & 47  & 12.9 & 96.1 & 9.0 & 9.0\\
18 & \textbf{8} & \textbf{44.9} & 97.6 & \textbf{44.9} & 17.4 & 45  & 43.5 & \textbf{97.9} & 43.5 & 17.4\\
*19 & \textbf{20} & \textbf{14.0} & 98.4 & \textbf{5.1} & 3.6 & 44  & 9.5 & \textbf{98.5} & 3.6 & 3.6\\
20 & \textbf{9} & \textbf{32.7} & 98.6 & \textbf{26.7} & \textbf{17.8} & 40  & 31.8 & 98.6 & 25.2 & 17.3\\
21 & \textbf{4} & 32.8 & \textbf{94.6} & 32.8 & 32.8 & 21  & 32.8 & 94.1 & 32.8 & 32.8\\
22 & \textbf{11} & 9.0 & \textbf{94.7} & \textbf{6.7} & \textbf{4.0} & 66  & 9.0 & 93.8 & 3.9 & 3.9\\
X & \textbf{18} & \textbf{50.6} & \textbf{98.6} & \textbf{27.1} & \textbf{13.2} & 64  & 40.1 & 98.3 & 11.7 & 11.7\\

    \bottomrule
    \end{tabular}
\end{table}

\begin{table}[h]
    \caption{Base-error metrics for our method (GatedGCN) and Raven's Layout heuristics on graphs of all the chromosomes, constructed from real reads. Asterisk indicates chromosome used in training.}
    \label{table:15xrealxerror}
    \small
    \centering
    \begin{tabular}{cC{1.8cm}C{1.5cm}C{1.5cm}C{1.5cm}}
    \toprule
        & \multicolumn{2}{c}{\textbf{GatedGCN}} & \multicolumn{2}{c}{\textbf{Raven}} \\
        \cmidrule(r){2-3}  \cmidrule(r){4-5}
    \textbf{chr} & \textbf{Mismatch} & \textbf{Indel} & \textbf{Mismatch} & \textbf{Indel} \\
    \midrule
    
1 & \textbf{2.54} & \textbf{0.91} & 5.30 & 1.21\\
2 & \textbf{1.50} & \textbf{0.64} & 2.23 & 0.85\\
3 & 3.47 & \textbf{0.69} & \textbf{2.46} & 0.73\\
4 & \textbf{1.32} & \textbf{0.65} & 3.63 & 0.75\\
5 & \textbf{2.65} & \textbf{0.54} & 4.20 & 0.74\\
6 & \textbf{0.84} & \textbf{0.50} & 1.09 & 0.56\\
7 & 2.89 & \textbf{0.99} & \textbf{2.29} & 1.16\\
8 & 2.53 & \textbf{0.72} & \textbf{1.79} & 0.73\\
9 & \textbf{5.22} & \textbf{1.94} & 8.98 & 2.59\\
10 & 3.60 & \textbf{0.93} & \textbf{2.85} & 1.12\\
11 & \textbf{0.65} & \textbf{0.74} & 1.59 & 1.04\\
12 & \textbf{0.37} & \textbf{0.53} & 1.68 & 0.67\\
13 & \textbf{2.16} & \textbf{0.63} & 8.95 & 1.50\\
14 & \textbf{1.57} & 1.18 & 1.80 & \textbf{1.14}\\
15 & \textbf{6.02} & \textbf{1.56} & 10.55 & 2.43\\
16 & \textbf{8.82} & \textbf{1.98} & 12.99 & 2.52\\
17 & \textbf{6.01} & \textbf{1.33} & 7.19 & 1.45\\
18 & \textbf{4.60} & \textbf{0.69} & 7.75 & 0.92\\
*19 & 9.45 & \textbf{1.84} & \textbf{8.48} & 2.09\\
20 & \textbf{4.69} & \textbf{0.93} & 9.05 & 1.65\\
21 & \textbf{4.46} & \textbf{1.52} & 10.00 & 1.82\\
22 & \textbf{24.42} & \textbf{2.32} & 27.45 & 4.40\\
X & \textbf{2.42} & \textbf{0.90} & 3.42 & 1.18\\

    \bottomrule
    \end{tabular}
    
\end{table}

\section{Discussion}
\label{section:discussion}

The experimental results clearly demonstrate practical value and usefulness of the proposed learning-based framework which opens new avenues for further research. Next steps will include application of the developed framework to genomes of different species and to different types of data, e.g. Oxford Nanopore Technology (ONT) reads. Since the assembly process with ONT reads is the same as with PacBio reads, we expect that similar improvements in the contiguity of the assemblies are possible. Furthermore, the generalization from one to all the other human chromosomes implies similar structure of chromosomes and we argue that generalizing to some other species (mainly mammals) is not far-fetched. One of the next challenges is haploid resolved assembly, which enables correct phasing of information from parental genomes. This, however, will require additional changes in the Overlap phase of genome assembly.

We are aware of other assemblers tailored for HiFi reads such as hifiasm \cite{cheng2021haplotype}, HiCanu \cite{nurk2020hicanu}, rust-mdbg \cite{ekim2021minimizer} and LJA \cite{bankevich2022multiplex}. However, the scope of this paper was not to present a whole new \textit{de novo} assembler, but to focus only on the Layout phase and to show that learning-based methods can bring an improvement over the conventional combination of algorithms and heuristics used in the field. For a fair comparison, there is a need for the same starting assembly graphs. Assemblers produce their own graphs which might significantly vary in type, complexity and fragmentation. Overlap step is a critical prerequisite for the successful reconstruction, but it is out of scope of this work. Having in mind the importance of Overlap step, presented framework is modular enough so the other researchers might incorporate it in other OLC based assemblers.

\section{Conclusion}
\label{section:conclusion}

In this work, we introduce a new framework for untangling large graphs constructed in \textit{de novo} assembly process with long reads. Existing \textit{de novo} assemblers use a combination of algorithms and hand-crafted heuristics which try to simplify graphs by removing some known structure found in them, and cut the graphs into fragments when heuristics cannot result in a unique solution. We propose a different approach, one based on graph neural networks, where we predict favorable and unfavorable edges for genome reconstruction, and then run a greedy search algorithm over these predictions. By starting from the graph produced by one of contemporary assemblers, Raven, we show that our method can improve the contiguity of the assemblies and nucleotide error rates. We trained our method only on the graphs generated from synthetic reads of a single chromosome, and are able to effectively untangle graphs generated from real human PacBio HiFi reads. We deem that presented framework proves usefulness of using graph neural networks in real life problems such as genome reconstruction.

\begin{ack}
We thank Robert Vaser for helping us with understanding the Overlap phase in Raven, as well as modifying it in order to work better for PacBio HiFi reads. We also thank Filip Bosnić and Sara Bakić for comments on the paper.

Lovro Vrček has been supported by "Young Researchers" Career Development Program DOK-2018-01-3373, ARAP scholarship awarded by A*STAR, and core funding of Genome Institute of Singapore, A*STAR.
Xavier Bresson has been supported by NRF Fellowship NRFF2017-10 and NUS-R-252-000-B97-133.
Martin Schmitz has been supported by SINGA scholarship awarded by A*STAR.
Mile Sikic has been supported in part by the European Union through the European Regional Development Fund under the grant KK.01.1.1.01.0009 (DATACROSS), by the Croatian Science Foundation under the project Single genome and metagenome assembly (IP-2018-01-5886), and by the core funding of Genome Institute of Singapore, A*STAR.

\end{ack}

\medskip

\bibliographystyle{unsrt}  
\bibliography{references}

\newpage

\appendix

\section{Code}
\label{supp:code}

Code can be found at \url{https://github.com/lvrcek/GNNome-assembly} and instructions for running it are in the \texttt{README.md} of the repository.

The main files in the repository are the following:
\begin{itemize}[leftmargin=*]
    \item \texttt{example.py} - Runs an entire framework on a small example, which includes setting up the directory structure, downloading real data, simulating the reads, constructing the graphs, training the model, and finding contigs in the assembly graph. Training is done on three chromosome 19 graphs, validation on one chromosome 19 graph, and inference on one chromosome 21 graph. Note that this, although a small example compared to other experiments, still takes time due to amount of data needed to download and generate.
    \item \texttt{config.py} - File where, inside three dictionaries, it can be specified graphs of which chromosome and in what amount should be used for training, validation, and testing.
    \item \texttt{pipeline.py} - Runs the entire framework, similar to the \texttt{example.py}, but generates and trains/validates/tests on the data specified in \texttt{config.py}.
    \item \texttt{reproduce.py} - Quickly reproduce the results reported in the paper, by running the script with \texttt{----mode} argument set to either \texttt{synth} for synthetic data or \texttt{real} for real data. Just like \texttt{example.py} and \texttt{pipeline.py}, it sets up the working directory in case it hasn't been set up before.
\end{itemize}

The easiest way to get started is to follow the installation instructions and then run \texttt{example.py} script. This will set up the directories for storing references and data. The default data-directory is \texttt{data/} and the default reference directory is \texttt{data/references/}. By running the code for the first time, CHM13 reference will be downloaded into \texttt{data/references/CHM13/}, in case it is not already located there. Real HiFi data, both the genomic sequences and the processed DGL graphs, will also be downloaded to \texttt{data/real/}. Compressed, the real dataset is 43 GB in size, while uncompressed it is 180 GB. Furthermore, by running the script, data-directory will be populated with directories \texttt{simulated/} and \texttt{experiments/}. All the simulated data for one of the chromosomes will be stored inside its respective directory inside \texttt{data/simulated/}. For example, simulated data for chromosome 15 will be stored inside \texttt{data/simulated/chr15/}.

These chromosome-specific directories follow a particular structure, in order to be compatible with \texttt{DGLDataset} class and to have all data related to one graph in one place:
\begin{itemize}[leftmargin=*]
    \item \texttt{raw/} - Directory where the genomic sequences are stored.
    \item \texttt{raven\_output/} - Directory where Raven stores assembly graphs after the Overlap phase and the assembly sequence after the Layout phase.
    \item \texttt{processed/} - Directory where the DGL graphs are stored.
    \item \texttt{info/} - Directory with auxiliary information about the graphs, used in training and inference.
    \item \texttt{graphia/} - Directory where a file in a format suitable for visualization in Graphia\footnote{Graphia is a tool for visualizing graphs. Read more about it here: \url{https://graphia.app/}} is stored.
\end{itemize}

For purposes of training and inference, simulated and real data used for training, validation, and testing is copied into \texttt{data/experiments/train\_<out>/}, \texttt{data/experiments/valid\_<out>/}, and \texttt{data/experiments/test\_<out>/}, respectively, where \texttt{<out>} is specified upon runtime and denotes the name of the run. Inference is performed only on the graphs in \texttt{data/experiments/test\_<out>/}, where two additional subdirectories are created:
\begin{itemize}[leftmargin=*]
    \item \texttt{inference/} - Directory where paths found during the decoding are stored.
    \item \texttt{assembly/} - Directory where the contigs, obtaiend by translating paths into sequences, are stored.
\end{itemize}

\newpage
\section{Model trained only on chromosome 19}
\label{supp:15xsynth}


The first model is trained on 15 graphs created from simulated chromosome 19 reads and is tested on graphs of all the chromosomes constructed from simulated (synthetic) data. Before comparing the model's performance against Raven's heuristics, we construct a sanity check by comparing it against two baseline methods---greedy search over overlap lengths and over overlap similarities. In order to make the comparison of the decoding with model-predicted probabilities and the baseline approaches fair, and avoid effects of the random sampling of edges, we first sample $B$ edges as starting positions from which we run three greedy searches, each following a different criteria. First baseline method chooses neighbors with the longest overlap, while the second baseline method chooses neighbors with the highest similarity. The results of these approaches can be seen in Table \ref{table:baselines}. The results of the decoding by following the edge-probabilities predicted by the model are shown in Table \ref{table:15xsynthxcontig}, together with the results produced by Raven. Note that in all three scenarios based on greedy search---following overlap lengths, similarities, and predicted probabilities---the number of contigs is the same, which is due to the sampling method described above. From this it is clear that both GatedGCN and Raven outperform na\"ive baseline approaches, indicating the complexity of the problem. It can also be noticed that GatedGCN outperforms Raven's heuristics by a significant margin. Moreover, in some cases NG50 and NGA50 were not computed, due to less than 50\% of the reference being reconstructed.

\begin{table}[h!]
    \caption{Contiguity metrics for the two baseline methods---greedy search choosing largest overlap length and greedy search choosing largest overlap similarity. Both baselines were run from the same sampled edges as greedy over model-provided probabilities, hence the same number of contigs.}
    \small
    \label{table:baselines}
    \centering
    \begin{tabular}{cC{0.9cm}C{1cm}C{0.7cm}C{0.7cm}C{1cm}C{0.9cm}C{1cm}C{0.7cm}C{0.7cm}C{1cm}}
    \toprule
        & \multicolumn{5}{c}{\textbf{Overlap length}} & \multicolumn{5}{c}{\textbf{Overlap similarity}} \\
        \cmidrule(r){2-6}  \cmidrule(r){7-11}
    \textbf{chr} & \textbf{Num ctg} & \textbf{Longest (Mbp)} & \textbf{GF (\%)} & \textbf{NG50 (Mbp)} & \textbf{NGA50 (Mbp)} & \textbf{Num ctg} &\textbf{Longest (Mbp)} & \textbf{GF (\%)} & \textbf{NG50 (Mbp)} & \textbf{NGA50 (Mbp)} \\
    \midrule

1 & 23 & 56.5 & 62.9 & 30.3 & 11.7 & 23  & 56.5 & 73.6 & 42.0 & 19.8\\
2 & 12 & 111.7 & 82.5 & 64.4 & 64.4 & 12  & 111.8 & 84.4 & 64.4 & 64.4\\
3 & 4 & 101.9 & 91.7 & 101.9 & 101.9 & 4  & 102.3 & 97.1 & 102.3 & 101.9\\
4 & 6 & 115.1 & 60.1 & 115.1 & 34.9 & 6  & 115.0 & 60.3 & 115.0 & 34.8\\
5 & 6 & 33.5 & 24.0 & - & - & 6  & 69.8 & 43.6 & - & - \\
6 & 5 & 26.2 & 34.9 & - & - & 5  & 32.0 & 38.5 & - & - \\
7 & 16 & 48.3 & 40.1 & - & - & 16  & 50.3 & 67.1 & 15.6 & 14.3\\
8 & 6 & 59.4 & 47.3 & - & - & 6  & 99.5 & 73.7 & 99.5 & 28.5\\
9 & 13 & 41.2 & 71.9 & 16.1 & 15.9 & 13  & 40.7 & 81.8 & 24.9 & 15.9\\
10 & 7 & 37.2 & 40.5 & - & - & 7  & 83.1 & 95.8 & 83.1 & 83.1\\
11 & 1 & 49.4 & 36.6 & - & - & 1  & 52.9 & 39.2 & - & - \\
12 & 4 & 57.5 & 96.4 & 34.9 & 34.8 & 4  & 59.1 & 97.4 & 34.9 & 34.6\\
13 & 8 & 96.1 & 88.3 & 96.1 & 57.2 & 8  & 96.1 & 92.9 & 96.1 & 57.2\\
14 & 9 & 9.8 & 15.0 & - & - & 9  & 86.8 & 95.1 & 86.8 & 86.8\\
15 & 14 & 47.3 & 70.3 & 13.4 & 13.4 & 14  & 41.0 & 75.2 & 13.4 & 13.4\\
16 & 21 & 18.7 & 52.7 & 0.6 & 0.1 & 21  & 18.5 & 64.0 & 10.0 & 2.0\\
17 & 8 & 19.2 & 56.2 & 7.4 & 2.2 & 8  & 28.3 & 57.4 & 17.8 & 12.2\\
18 & 3 & 59.1 & 94.3 & 59.1 & 26.1 & 3  & 59.2 & 94.5 & 59.2 & 26.5\\
19 & 2 & 31.7 & 51.5 & 31.7 & 0.9 & 2  & 31.6 & 51.5 & 31.6 & 0.9\\
20 & 4 & 36.9 & 94.5 & 36.9 & 33.9 & 4  & 37.1 & 94.7 & 37.1 & 33.4\\
21 & 1 & 4.3 & 9.5 & - & - & 1  & 8.9 & 19.2 & - & - \\
22 & 4 & 26.8 & 62.4 & 26.8 & 26.8 & 4  & 26.9 & 72.0 & 26.9 & 26.9\\
X & 9 & 17.5 & 32.3 & - & - & 9  & 46.9 & 91.1 & 39.4 & 39.2\\

    \bottomrule
    \end{tabular}
    
\end{table}

    

    

\begin{table}[h!]
    \caption{Contiguity metrics for our method (GatedGCN) and Raven's Layout heuristics on graphs of all the chromosomes, constructed from simulated reads. Asterisk indicates chromosome used in training, although the graph itself was different.}
    \label{table:15xsynthxcontig}
    \small
    \centering
    \begin{tabular}{cC{0.9cm}C{1cm}C{0.7cm}C{0.7cm}C{1cm}C{0.9cm}C{1cm}C{0.7cm}C{0.7cm}C{1cm}}
    \toprule
        & \multicolumn{5}{c}{\textbf{GatedGCN}} & \multicolumn{5}{c}{\textbf{Raven}} \\
        \cmidrule(r){2-6}  \cmidrule(r){7-11}
    \textbf{chr} & \textbf{Num ctg} & \textbf{Longest (Mbp)} & \textbf{GF (\%)} & \textbf{NG50 (Mbp)} & \textbf{NGA50 (Mbp)} & \textbf{Num ctg} &\textbf{Longest (Mbp)} & \textbf{GF (\%)} & \textbf{NG50 (Mbp)} & \textbf{NGA50 (Mbp)} \\
    \midrule

1 & \textbf{23} & \textbf{77.9} & \textbf{97.8} & \textbf{56.5} & \textbf{30.5} & 233  & 56.4 & 97.6 & 30.4 & 30.4\\
2 & \textbf{12} & \textbf{111.8} & \textbf{99.2} & \textbf{87.8} & \textbf{87.0} & 57  & 111.0 & 99.1 & 86.7 & 86.7\\
3 & \textbf{4} & \textbf{102.8} & 99.2 & \textbf{102.8} & \textbf{101.9} & 36  & 101.8 & \textbf{99.4} & 101.8 & 101.8\\
4 & \textbf{6} & \textbf{188.7} & 99.0 & \textbf{188.7} & 58.4 & 60  & 67.5 & \textbf{99.1} & 58.4 & 58.4\\
5 & \textbf{6} & \textbf{169.6} & 99.1 & \textbf{169.6} & \textbf{101.2} & 51  & 101.1 & 99.1 & 101.1 & 101.1\\
6 & \textbf{5} & \textbf{133.0} & \textbf{98.8} & \textbf{133.0} & \textbf{100.9} & 18  & 110.4 & 98.6 & 110.4 & 100.3\\
7 & \textbf{16} & \textbf{69.2 }& \textbf{98.2} & \textbf{54.4} & \textbf{25.7} & 57  & 40.9 & 98.0 & 26.7 & 24.1\\
8 & \textbf{6} & \textbf{99.5} & \textbf{98.6} & \textbf{99.5} & \textbf{32.0} & 27  & 59.3 & 98.4 & 31.9 & 31.9\\
9 & \textbf{13} & 41.2 & \textbf{91.4} & \textbf{24.9} & 15.9 & 133  & 41.2 & 89.9 & 16.1 & 15.9\\
10 & \textbf{7} & \textbf{83.2} & \textbf{99.3} & \textbf{83.2} & \textbf{83.1} & 42  & 46.4 & 99.1 & 37.2 & 37.2\\
11 & \textbf{1} & \textbf{134.9} & \textbf{99.8} & \textbf{134.9} & \textbf{32.9} & 32  & 40.9 & 99.7 & 31.3 & 31.3\\
12 & \textbf{4} & 59.1 & 99.7 & \textbf{37.7} & 34.8 & 21  & 59.1 & 99.7 & 35.0 & 34.8\\
13 & \textbf{8} & \textbf{96.6} & \textbf{96.2} & \textbf{96.6} & 57.2 & 98  & 62.0 & 95.7 & 62.0 & 57.2\\
14 & \textbf{9} & \textbf{86.8} & \textbf{97.8} & \textbf{86.8} & \textbf{86.8} & 30  & 86.0 & 97.0 & 86.0 & 86.0\\
15 & \textbf{14} & \textbf{41.0} & \textbf{93.8} & \textbf{13.4} & \textbf{13.4} & 134  & 29.1 & 92.5 & 9.0 & 9.0\\
16 & \textbf{21} & \textbf{34.5} & \textbf{91.6} & \textbf{21.3} & \textbf{15.8} & 136  & 15.9 & 89.6 & 9.5 & 9.5\\
17 & \textbf{8} & \textbf{24.3} & \textbf{96.7} & \textbf{19.8} & \textbf{13.2} & 41  & 18.3 & 96.2 & 12.9 & 12.9\\
18 & \textbf{3} & \textbf{59.9} & 97.4 & \textbf{59.9} & \textbf{26.5} & 34  & 59.4 & \textbf{97.5} & 59.4 & 26.0\\
*19 & \textbf{2} & \textbf{60.7} & \textbf{98.6} & \textbf{60.7} & 9.5 & 22  & 16.0 & 98.1 & 9.5 & 9.5\\
20 & \textbf{4} & \textbf{37.1} & \textbf{98.5} & \textbf{37.1} & \textbf{34.3} & 36  & 33.9 & 98.4 & 33.9 & 33.9\\
21 & \textbf{1} & \textbf{42.4} & 94.1 & \textbf{42.4} & 33.8 & 25  & 33.8 & \textbf{94.2} & 33.8 & 33.8\\
22 & \textbf{4} & 26.9 & \textbf{95.1} & 26.9 & 26.9 & 54  & 26.9 & 94.0 & 26.9 & 26.9\\
23 & \textbf{9} & \textbf{51.4} & \textbf{98.6} & \textbf{39.4} & \textbf{39.2} & 53  & 47.8 & 98.2 & 15.8 & 14.2\\

    \bottomrule
    \end{tabular}
    
\end{table}

We also compare the per-base errors---number of mismatches and indels (insertions and deletions) per 100,000 base pairs---of assemblies produced by GatedGCN and Raven. We can notice that the error rate between the chromosomes varies greatly, but, when compared side-by-side, it is clear that GatedGCN outperforms Raven for most of the chromosomes. This can be seen in Table \ref{table:15xsynthxerror}.

\begin{table}[H]
    \caption{Base-error metrics for our method (GatedGCN) and Raven's Layout heuristics on graphs of all the chromosomes, constructed from simulated reads. Asterisk indicates chromosome used in training, although the graph itself was different.}
    \label{table:15xsynthxerror}
    \small
    \centering
    \begin{tabular}{cC{1.8cm}C{1.5cm}C{1.5cm}C{1.5cm}}
    \toprule
        & \multicolumn{2}{c}{\textbf{GatedGCN}} & \multicolumn{2}{c}{\textbf{Raven}} \\
        \cmidrule(r){2-3}  \cmidrule(r){4-5}
    \textbf{chr} & \textbf{Mismatch} & \textbf{Indel} & \textbf{Mismatch} & \textbf{Indel} \\
    \midrule
    
1 & \textbf{4.41} & \textbf{0.46} & 6.21 & 0.52\\
2 & \textbf{0.72} & \textbf{0.10} & 1.54 & 0.26\\
3 & 2.83 & 0.09 & \textbf{2.16} & 0.09\\
4 & \textbf{0.84} & \textbf{0.04} & 3.40 & 0.22\\
5 & 5.04 & \textbf{0.17} & \textbf{4.09} & 0.33\\
6 & 1.04 & \textbf{0.05} & \textbf{0.61} & 0.06\\
7 & \textbf{1.02} & \textbf{0.11} & 3.09 & 0.40\\
8 & \textbf{1.25} & \textbf{0.16} & 2.49 & 0.18\\
9 & \textbf{5.16} & \textbf{0.88} & 11.76 & 2.04\\
10 & 7.17 & 0.39 & \textbf{1.86} & \textbf{0.20}\\
11 & \textbf{0.42} & \textbf{0.06} & 0.74 & 0.07\\
12 & \textbf{0.57} & \textbf{0.01} & 1.05 & 0.05\\
13 & \textbf{1.24} & \textbf{0.08} & 7.31 & 0.54\\
14 & 2.07 & 0.24 & \textbf{1.64} & \textbf{0.17}\\
15 & \textbf{5.08} & \textbf{0.89} & 9.91 & 1.5\\
16 & \textbf{7.78} & \textbf{0.92} & 14.04 & 1.59\\
17 & \textbf{6.15} & \textbf{0.42} & 6.21 & 0.52\\
18 & \textbf{4.34} & \textbf{0.17} & 4.89 & 0.19\\
*19 & 6.91 & 0.40 & \textbf{3.89} & \textbf{0.12}\\
20 & \textbf{7.54} & \textbf{0.26} & 10.45 & 0.73\\
21 & \textbf{2.71} & \textbf{0.19} & 4.93 & 0.54\\
22 & \textbf{24.59} & \textbf{1.01} & 29.14 & 1.43\\
23 & \textbf{2.77} & \textbf{0.21} & 3.15 & 0.22\\

    \bottomrule
    \end{tabular}
    
\end{table}

\newpage
\section{Model trained on chromosomes 9, 19, and 22}
\label{supp:mix-model}

In Section 4.1 of the paper, we mention that training only on chromosome 19 is not the only experiment we performed, but also tried to train on a combination of graphs from chromosomes 9, 19, and 22. It is expected that training on 15 graphs of a combination of different chromosomes would improve performance over training on 15 graph of a single chromosome. This is indeed the case here as well, also proving that there is nothing special about chromosome 19. The model trained on the same amount of graphs of different chromosomes does perform slightly better, but the improvements are not statistically significant. Therefore, in the paper we report only the results obtained by training on a single chromosome.

We test the model trained on chromosomes 9, 19, and 22, on both synthetic and real data, and compare both the contiguity and per-base errors against Raven. Contiguity on synthetic data can be seen in Table \ref{table:mixsynthcontiguity}, per-base errors on synthetic data can be seen in Table \ref{table:mixsyntherrors}, contiguity on real data can be seen in Table \ref{table:mixrealcontiguity}, per-base errors on real data can be seen in Table \ref{table:mixrealerrors}.


\begin{table}[H]
    \caption{Contiguity metrics for our method (GatedGCN), trained on three chromosomes, and Raven's Layout heuristics on graphs of all the chromosomes, constructed from simulated reads. Asterisk indicates chromosomes used in training, although the graphs were different.}
    \small
    \label{table:mixsynthcontiguity}
    \centering
    \begin{tabular}{cC{0.9cm}C{1cm}C{0.7cm}C{0.7cm}C{1cm}C{0.9cm}C{1cm}C{0.7cm}C{0.7cm}C{1cm}}
    \toprule
        & \multicolumn{5}{c}{\textbf{GatedGCN}} & \multicolumn{5}{c}{\textbf{Raven}} \\
        \cmidrule(r){2-6}  \cmidrule(r){7-11}
    \textbf{chr} & \textbf{Num ctg} & \textbf{Longest (Mbp)} & \textbf{GF (\%)} & \textbf{NG50 (Mbp)} & \textbf{NGA50 (Mbp)} & \textbf{Num ctg} &\textbf{Longest (Mbp)} & \textbf{GF (\%)} & \textbf{NG50 (Mbp)} & \textbf{NGA50 (Mbp)} \\
    \midrule

1 & \textbf{27} & \textbf{135.0} & \textbf{97.8} & \textbf{135.0} & \textbf{30.5} & 233  & 56.4 & 97.6 & 30.4 & 30.4\\
2 & \textbf{13} & \textbf{111.8} & \textbf{99.2} & \textbf{87.8} & \textbf{87.0} & 57  & 111.0 & 99.1 & 86.7 & 86.7\\
3 & \textbf{4} & \textbf{102.7} & 99.1 & \textbf{102.7} & \textbf{101.9} & 36  & 101.8 & \textbf{99.4} & 101.8 & 101.8\\
4 & \textbf{6} & \textbf{139.0} & 98.9 & \textbf{139.0} & 58.3 & 60  & 67.5 & \textbf{99.1} & 58.4 & \textbf{58.4}\\
5 & \textbf{8} & \textbf{121.5} & \textbf{99.3} & \textbf{121.5} & \textbf{101.2} & 51  & 101.1 & 99.1 & 101.1 & 101.1\\
6 & \textbf{5} & 111.0 & \textbf{98.7} & 111.0 & \textbf{100.9} & 18  & \textbf{110.4} & 98.6 & \textbf{110.4} & 100.3\\
7 & \textbf{15} & \textbf{67.2} & \textbf{98.2} & \textbf{54.4} & \textbf{41.5} & 57  & 40.9 & 98.0 & 26.7 & 24.1\\
8 & \textbf{6} & \textbf{99.6} & \textbf{98.8} & \textbf{99.6} & 31.4 & 27  & 59.3 & 98.4 & 31.9 & \textbf{31.9} \\
*9 & \textbf{17} & 41.2 & \textbf{92.4} & \textbf{17.6} & 15.9 & 133  & 41.2 & 89.9 & 16.1 & 15.9\\
10 & \textbf{7} & \textbf{83.2} & \textbf{99.3} & \textbf{83.2} & \textbf{83.1} & 42  & 46.4 & 99.1 & 37.2 & 37.2\\
11 & \textbf{1} & \textbf{134.9} & \textbf{99.8} & \textbf{134.9} & \textbf{35.3} & 32  & 40.9 & 99.7 & 31.3 & 31.3\\
12 & \textbf{5} & 59.1 & \textbf{99.8} & \textbf{37.6} & 34.8 & 21  & 59.1 & 99.7 & 35.0 & 34.8\\
13 & \textbf{7} & \textbf{96.2} & \textbf{95.9} & \textbf{96.2} & 57.2 & 98  & 62.0 & 95.7 & 62.0 & 57.2\\
14 & \textbf{8} & \textbf{86.8} & \textbf{97.8} & \textbf{86.8} & \textbf{86.8} & 30  & 86.0 & 97.0 & 86.0 & 86.0\\
15 & \textbf{20} & \textbf{41.0} & \textbf{94.3} & \textbf{13.4} & \textbf{13.4} & 134  & 29.1 & 92.5 & 9.0 & 9.0\\
16 & \textbf{19} & \textbf{34.5} & \textbf{92.0} & \textbf{21.3} & \textbf{15.8} & 136  & 15.9 & 89.6 & 9.5 & 9.5\\
17 & \textbf{7} & \textbf{38.7} & \textbf{96.6} & \textbf{20.5} & \textbf{18.4} & 41  & 18.3 & 96.2 & 12.9 & 12.9\\
18 & \textbf{4} & \textbf{59.9} & 97.3 & \textbf{59.9} & \textbf{26.5} & 34  & 59.4 & \textbf{97.5} & 59.4 & 26.0\\
*19 & \textbf{3} & \textbf{55.7} & \textbf{98.5} & \textbf{55.7} & 9.5 & 22  & 16.0 & 98.1 & 9.5 & 9.5\\
20 & \textbf{3} & \textbf{37.1} & 98.2 & \textbf{37.1} & \textbf{34.3} & 36  & 33.9 & \textbf{98.4} & 33.9 & 33.9\\
21 & \textbf{4} & \textbf{38.9} & \textbf{94.6} & \textbf{38.9} & 30.4 & 25  & 33.8 & 94.2 & 33.8 & \textbf{33.8}\\
*22 & \textbf{6} & 26.9 & \textbf{94.5} & 26.9 & 26.9 & 54  & 26.9 & 94.0 & 26.9 & 26.9\\
X & \textbf{6} & \textbf{51.4} & \textbf{98.5} & \textbf{47.1} & \textbf{40.0} & 53  & 47.8 & 98.2 & 15.8 & 14.2\\

    \bottomrule
    \end{tabular}
    
\end{table}

\begin{table}[H]
    \caption{Base-error metrics for our method (GatedGCN), trained on three chromosomes, and Raven's Layout heuristics on graphs of all the chromosomes, constructed from simulated reads. Asterisk indicates chromosomes used in training, although the graphs were different.}
    \label{table:mixsyntherrors}
    \small
    \centering
    \begin{tabular}{cC{1.5cm}C{1.5cm}C{1.5cm}C{1.5cm}}
    \toprule
        & \multicolumn{2}{c}{\textbf{GatedGCN}} & \multicolumn{2}{c}{\textbf{Raven}} \\
        \cmidrule(r){2-3}  \cmidrule(r){4-5}
    \textbf{chr} & \textbf{Mismatch} & \textbf{Indel} & \textbf{Mismatch} & \textbf{Indel} \\
    \midrule
    
1 & \textbf{3.00} & \textbf{0.22} & 6.21 & 0.52\\
2 & \textbf{0.96} & \textbf{0.11} & 1.54 & 0.26\\
3 & 2.89 & \textbf{0.08} & \textbf{2.16} & 0.09\\
4 & \textbf{1.05} & \textbf{0.06} & 3.40 & 0.22\\
5 & \textbf{3.76} & \textbf{0.16} & 4.09 & 0.33\\
6 & 0.86 & \textbf{0.04} & \textbf{0.61} & 0.06\\
7 & \textbf{1.22} & \textbf{0.11} & 3.09 & 0.40\\
8 & \textbf{1.43} & 0.18 & 2.49 & 0.18\\
*9 & \textbf{6.19} & \textbf{1.10} & 11.76 & 2.04\\
10 & 4.11 & 0.27 & \textbf{1.86} & \textbf{0.20}\\
11 & \textbf{0.32} & \textbf{0.04} & 0.74 & 0.07\\
12 & \textbf{0.82} & \textbf{0.02} & 1.05 & 0.05\\
13 & \textbf{2.75} & \textbf{0.13} & 7.31 & 0.54\\
14 & \textbf{0.91} & \textbf{0.15} & 1.64 & 0.17\\
15 & \textbf{4.68} & \textbf{0.60} & 9.91 & 1.50\\
16 & \textbf{7.38} & \textbf{0.82} & 14.04 & 1.59\\
17 & 6.61 & 0.60 & \textbf{6.21} & \textbf{0.52}\\
18 & \textbf{4.38} & \textbf{0.09} & 4.89 & 0.19\\
*19 & 8.40 & 0.53 & \textbf{3.89} &\textbf{0.12}\\
20 & \textbf{6.25} & \textbf{0.20} & 10.45 & 0.73\\
21 & \textbf{3.20} & \textbf{0.19} & 4.93 & 0.54\\
*22 & \textbf{21.92} & \textbf{0.92} & 29.14 & 1.43\\
X & \textbf{3.09} & \textbf{0.18} & 3.15 & 0.22\\

    \bottomrule
    \end{tabular}
    
\end{table}


\begin{table}[H]
    \caption{Contiguity metrics for our method (GatedGCN), trained on three chromosomes, and Raven's Layout heuristics on graphs of all the chromosomes, constructed from real reads. Asterisk indicates chromosomes used in training.}
    \small
    \label{table:mixrealcontiguity}
    \centering
    \begin{tabular}{cC{0.9cm}C{1cm}C{0.7cm}C{0.7cm}C{1cm}C{0.9cm}C{1cm}C{0.7cm}C{0.7cm}C{1cm}}
    \toprule
        & \multicolumn{5}{c}{\textbf{GatedGCN}} & \multicolumn{5}{c}{\textbf{Raven}} \\
        \cmidrule(r){2-6}  \cmidrule(r){7-11}
    \textbf{chr} & \textbf{Num ctg} & \textbf{Longest (Mbp)} & \textbf{GF (\%)} & \textbf{NG50 (Mbp)} & \textbf{NGA50 (Mbp)} & \textbf{Num ctg} &\textbf{Longest (Mbp)} & \textbf{GF (\%)} & \textbf{NG50 (Mbp)} & \textbf{NGA50 (Mbp)} \\
    \midrule

1 & \textbf{33} & \textbf{103.7} & \textbf{98.4} & \textbf{63.6} & \textbf{46.3} & 241  & 86.9 & 97.6 & 44.4 & 44.4\\
2 & \textbf{19} & 73.1 & \textbf{99.5} & \textbf{35.1} & \textbf{28.2} & 56  & 73.1 & 98.9 & 28.1 & 28.1\\
3 & \textbf{8} & 90.5 & 99.5 & 56.0 & 56.0 & 45  & 90.5 & 99.5 & 56.0 & 56.0\\
4 & \textbf{12} & \textbf{115.2} & 99.0 & \textbf{115.2} & 34.8 & 78  & 67.8 & 99.0 & 34.9 & \textbf{34.9} \\
5 & \textbf{8} & \textbf{123.6} & \textbf{99.1} & \textbf{123.6} & 103.5 & 47  & 103.5 & 99.0 & 103.5 & 103.5\\
6 & \textbf{7} & 101.0 & \textbf{98.8} & 101.0 & \textbf{52.8} & 20  & \textbf{110.3} & 98.7 & \textbf{110.3} & 25.9\\
7 & \textbf{19} & \textbf{58.7} & \textbf{98.4} & \textbf{42.6} & \textbf{25.7} & 69  & 29.3 & 98.0 & 25.1 & 17.5\\
8 & \textbf{14} & \textbf{68.7} & 98.4 & \textbf{34.9} & \textbf{28.5} & 33  & 31.6 & 98.4 & 28.5 & 28.5\\
*9 & \textbf{19} & \textbf{66.8} & \textbf{90.9} & \textbf{31.9} & \textbf{16.1} & 139  & 38.9 & 90.2 & 19.7 & 15.8\\
10 & \textbf{14} & \textbf{40.1} & 99.2 & \textbf{36.7} & \textbf{31.9} & 43  & 36.7 & 99.2 & 17.2 & 17.2\\
11 & \textbf{7} & \textbf{66.2} & \textbf{99.8} & \textbf{35.3} & \textbf{32.8} & 31  & 35.3 & 99.7 & 32.6 & 23.2\\
12 & \textbf{12} & 57.2 & \textbf{99.9} & 31.0 & 31.0 & 33  & 57.2 & 99.8 & 31.0 & 31.0\\
13 & \textbf{15} & \textbf{73.0} & \textbf{96.1} & \textbf{73.0} & \textbf{30.1} & 116  & 47.5 & 95.9 & 25.5 & 25.5\\
14 & \textbf{9} & 82.6 & \textbf{97.5} & 82.6 & 82.6 & 32  & 82.6 & 97.2 & 82.6 & 82.6\\
15 & \textbf{24} & \textbf{47.0} & \textbf{93.9} & \textbf{13.4} & \textbf{10.0} & 157  & 29.0 & 93.5 & 9.0 & 8.5\\
16 & \textbf{29} & 16.0 & \textbf{90.9} & \textbf{8.7} & \textbf{8.7} & 164  & \textbf{16.4} & 90.8 & 5.9 & 5.7\\
17 & \textbf{13} & \textbf{22.7} & \textbf{96.4} & \textbf{10.8} & \textbf{10.2} & 47  & 12.9 & 96.1 & 9.0 & 9.0\\
18 & \textbf{6} & \textbf{45.0} & 97.4 & \textbf{45.0} & 17.4 & 45  & 43.5 & \textbf{97.9} & 43.5 & 17.4\\
*19 & \textbf{19} & \textbf{12.4} & 98.5 & \textbf{5.1} & 3.6 & 44  & 9.5 & 98.5 & 3.6 & 3.6\\
20 & \textbf{10} & \textbf{32.7} & \textbf{98.7} & \textbf{26.6} & \textbf{26.1} & 40  & 31.8 & 98.6 & 25.2 & 17.3\\
21 & \textbf{4} & 32.8 & \textbf{94.5} & 32.8 & 32.8 & 21  & 32.8 & 94.1 & 32.8 & 32.8\\
*22 & \textbf{11} & \textbf{11.0} & \textbf{95.3} & \textbf{8.3} & \textbf{4.1} & 66  & 9.0 & 93.8 & 3.9 & 3.9\\
X & \textbf{17} & \textbf{50.6} & \textbf{98.5} & \textbf{27.1} & \textbf{13.2} & 64  & 40.1 & 98.3 & 11.7 & 11.7\\

    \bottomrule
    \end{tabular}
    
\end{table}

\begin{table}[H]
    \caption{Base-error metrics for our method (GatedGCN), trained on three chromosomes, and Raven's Layout heuristics on graphs of all the chromosomes, constructed from real reads. Asterisk indicates chromosomes used in training.}
    \label{table:mixrealerrors}
    \small
    \centering
    \begin{tabular}{cC{1.5cm}C{1.5cm}C{1.5cm}C{1.5cm}}
    \toprule
        & \multicolumn{2}{c}{\textbf{GatedGCN}} & \multicolumn{2}{c}{\textbf{Raven}} \\
        \cmidrule(r){2-3}  \cmidrule(r){4-5}
    \textbf{chr} & \textbf{Mismatch} & \textbf{Indel} & \textbf{Mismatch} & \textbf{Indel} \\
    \midrule
    
1 & \textbf{1.95} & \textbf{0.79} & 5.30 & 1.21\\
2 & \textbf{0.81} & \textbf{0.58} & 2.23 & 0.85\\
3 & 2.90 & \textbf{0.65} & \textbf{2.46} & 0.73\\
4 & \textbf{1.21} & \textbf{0.72} & 3.63 & 0.75\\
5 & \textbf{2.85} & \textbf{0.54} & 4.20 & 0.74\\
6 & \textbf{0.65} & \textbf{0.45} & 1.09 & 0.56\\
7 & 2.36 & \textbf{0.84} & \textbf{2.29} & 1.16\\
8 & 1.87 & \textbf{0.67} & \textbf{1.79} & 0.73\\
*9 & \textbf{4.17} & \textbf{1.58} & 8.98 & 2.59\\
10 & 7.06 & \textbf{1.08} & \textbf{2.85} & 1.12\\
11 & \textbf{0.78} & \textbf{0.78} & 1.59 & 1.04\\
12 & \textbf{0.68} & \textbf{0.56} & 1.68 & 0.67\\
13 & \textbf{2.63} & \textbf{0.58} & 8.95 & 1.50\\
14 & \textbf{1.27} & \textbf{0.97} & 1.80 & 1.14\\
15 & \textbf{7.80} & \textbf{1.99} & 10.55 & 2.43\\
16 & \textbf{7.55} & \textbf{1.94} & 12.99 & 2.52\\
17 & \textbf{6.62} & 1.57 & 7.19 & \textbf{1.45}\\
18 & \textbf{6.10} & \textbf{0.68} & 7.75 & 0.92\\
*19 & 9.66 & \textbf{2.03} & \textbf{8.48} & 2.09\\
20 & \textbf{3.34} & \textbf{1.05} & 9.05 & 1.65\\
21 & \textbf{3.83} & 1.95 & 10.00 & \textbf{1.82}\\
*22 & \textbf{26.99} & \textbf{2.98} & 27.45 & 4.40\\
X & \textbf{2.16} & \textbf{0.84} & 3.42 & 1.18\\

    \bottomrule
    \end{tabular}
    
\end{table}

\newpage

\section{Dataset}
\label{supp:dataset}

The human reference genome, CHM13, consists of 23 chromosomes, which are together 3.3 billion base pairs long. The PacBio HiFi dataset from which CHM13 was reconstructed, and which we use to evaluate our model, consists of 5.6 million reads. The amount of base pairs and reads per each chromosome can be seen in Table \ref{table:chm13-stat}, together with the number of nodes and edges in each graph corresponding to a particular chromosome. During the graph construction, some of the reads are found to be contained inside other, longer reads, which is why the graphs have less nodes than the chromosomes have reads.

\begin{table}[h!]
    \centering
    \caption{Statistics for data set based on real HiFi reads, aligned with minimap2 and assembly graph created with Raven }
    \label{table:chm13-stat}
    \begin{tabular}{cC{1.5cm}C{1.5cm}C{1.5cm}C{1.5cm}}
    \toprule
    \textbf{chr} & \textbf{Base pairs} & \textbf{Reads} & \textbf{Nodes} & \textbf{Edges} \\
    \midrule
    
1 & 248,387,328 & 462,582 & 184,050 & 1,407,158 \\
2 & 242,696,752 & 444,450 & 180,764 & 1,381,376 \\
3 & 201,105,948 & 366,547 & 149,828 & 1,138,668 \\
4 & 193,574,945 & 352,056 & 145,702 & 1,115,808 \\
5 & 182,045,439 & 332,985 & 135,068 & 1,028,484 \\
6 & 172,126,628 & 313,731 & 127,698 & 966,560 \\
7 & 160,567,428 & 291,366 & 120,280 & 890,388 \\
8 & 146,259,331 & 265,288 & 108,948 & 828,772 \\
9 & 150,617,247 & 290,786 & 106,874 & 860,710 \\
10 & 134,758,134 & 244,927 & 101,484 & 765,180 \\
11 & 135,127,769 & 246,436 & 100,598 & 757,642 \\
12 & 133,324,548 & 241,403 & 99,542 & 750,364 \\
13 & 113,566,686 & 199,405 & 84,500 & 653,730 \\
14 & 101,161,492 & 182,551 & 73,436 & 541,214 \\
15 & 99,753,195 & 183,176 & 70,598 & 535,842 \\
16 & 96,330,374 & 182,280 & 65,834 & 519,358 \\
17 & 84,276,897 & 150,066 & 60,498 & 439,416 \\
18 & 80,542,538 & 147,509 & 59,868 & 459,356 \\
19 & 61,707,364 & 105,052 & 45,114 & 315,348 \\
20 & 66,210,255 & 120,635 & 48,816 & 366,614 \\
21 & 45,090,682 & 79,245 & 32,096 & 239,166 \\
22 & 51,324,926 & 89,624 & 35,612 & 252,666 \\
X & 154,259,566 & 272,496 & 112,922 & 834,702 \\
    
    \bottomrule
    \end{tabular}
    
\end{table}

\begin{figure}
  \centering
  
      \includegraphics[width=0.6\linewidth]{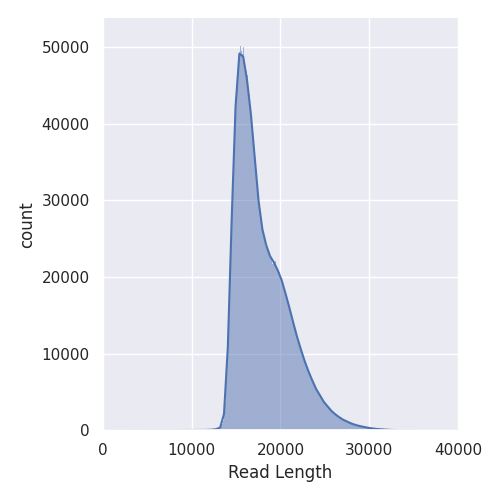}
  \caption{Histogram of read length of real HiFi reads from the dataset (the synthetic reads have exactly the same length distribution).}
  \label{fig:readlength}
\end{figure}

\begin{figure}
  \centering
  
      \includegraphics[width=0.6\linewidth]{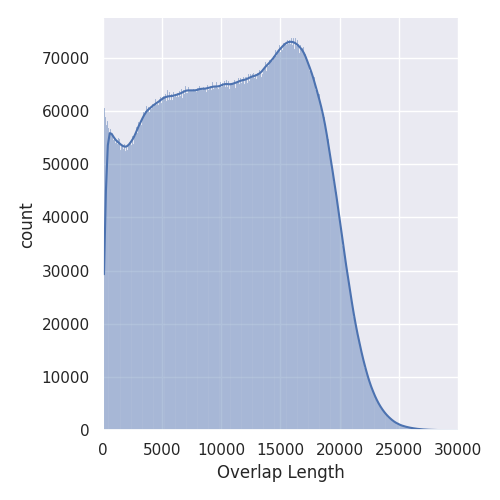}
  \caption{Histogram of the overlap length of the Raven assembly graph based on the real HiFi reads from the dataset.}
  \label{fig:overlaplength}
\end{figure}

\begin{figure}
\begin{tabular}{cc}
  \includegraphics[width=0.5\linewidth]{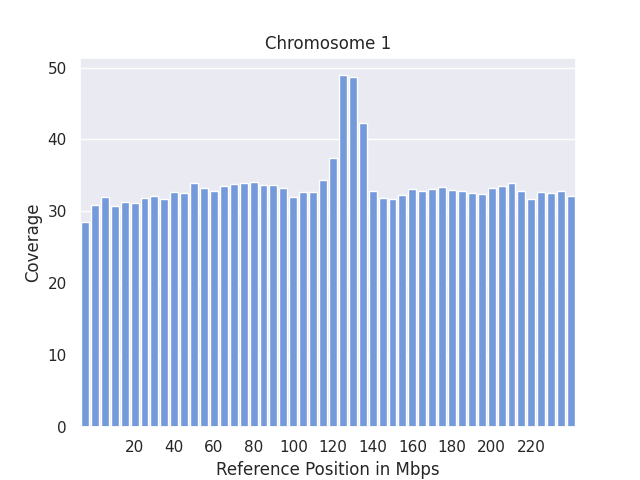} &   \includegraphics[width=0.5\linewidth]{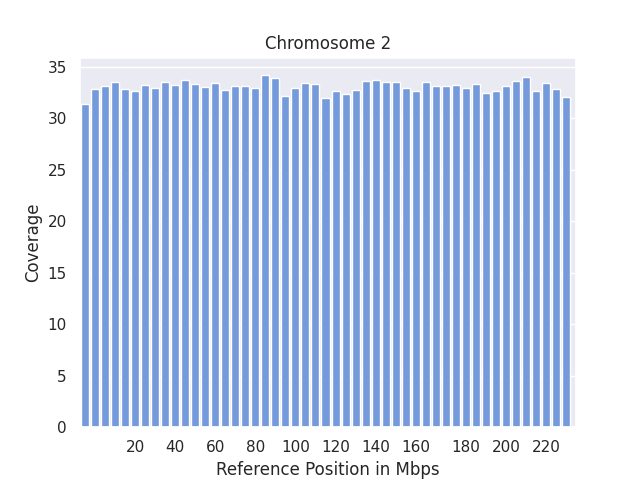} \\
  \includegraphics[width=0.5\linewidth]{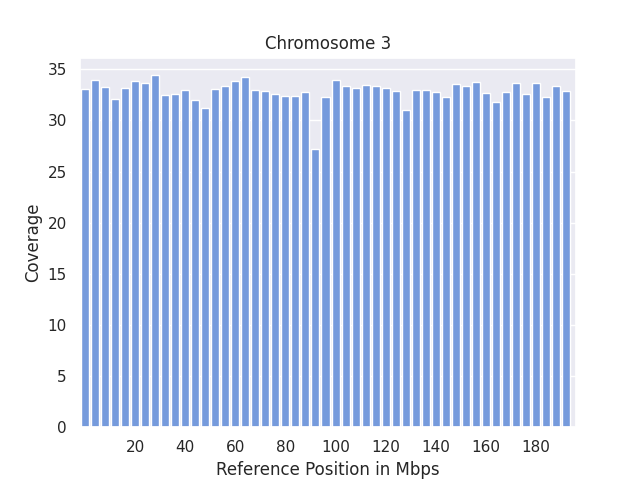} &  \includegraphics[width=0.5\linewidth]{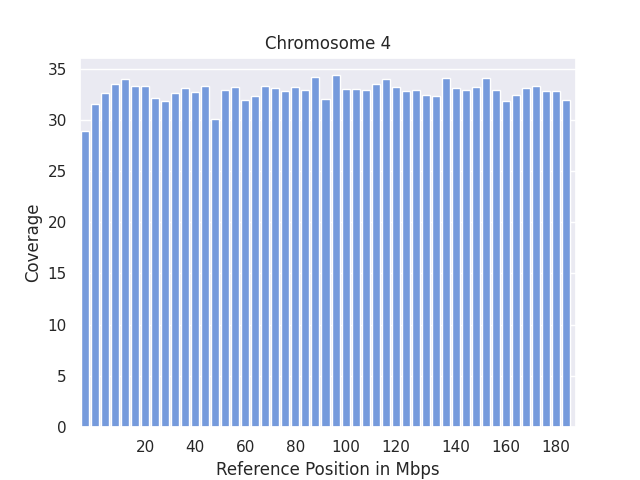} \\
  \includegraphics[width=0.5\linewidth]{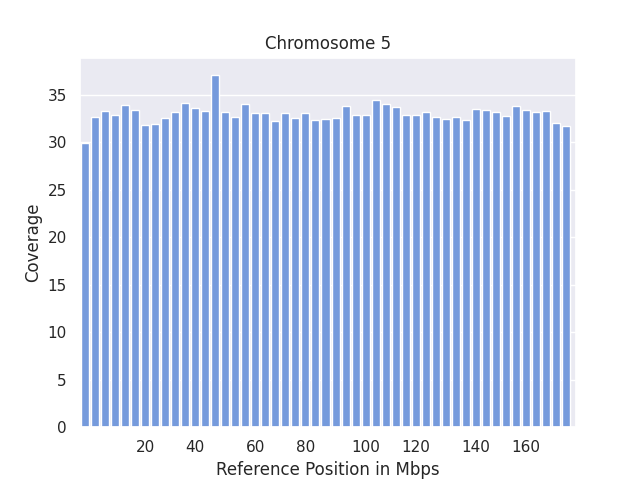} &
  \includegraphics[width=0.5\linewidth]{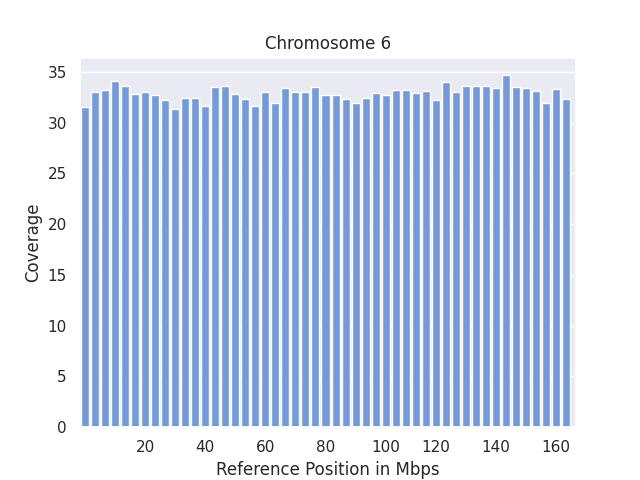} \\
  \includegraphics[width=0.5\linewidth]{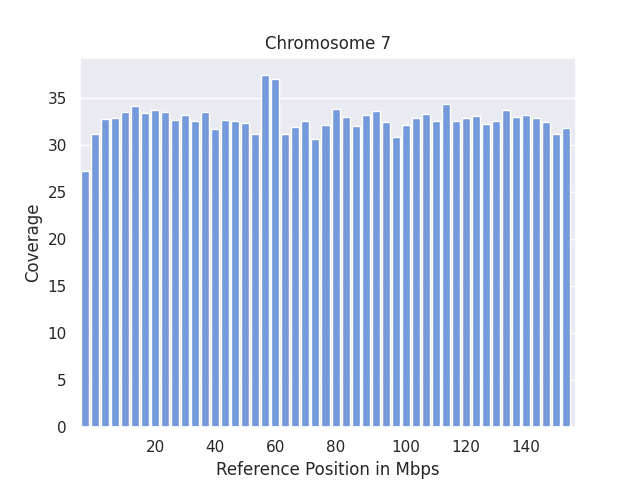} &
  \includegraphics[width=0.5\linewidth]{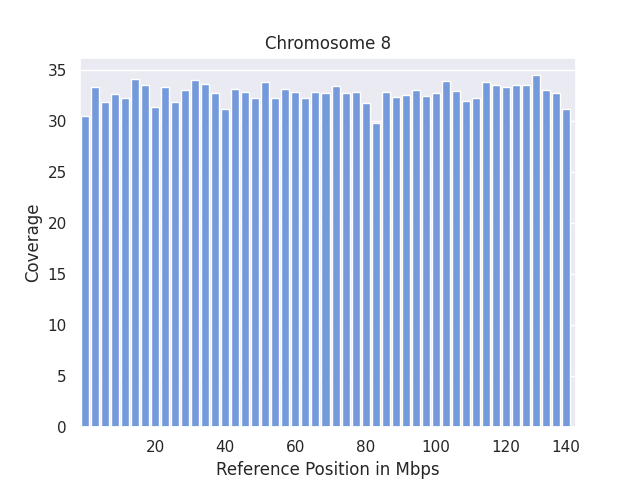} \\
\end{tabular}
\caption{Coverage histograms for real HiFi reads of chromosomes 1 to 8.}
\label{fig:cov1}
\end{figure}

\begin{figure}
\begin{tabular}{cc}

  \includegraphics[width=0.5\linewidth]{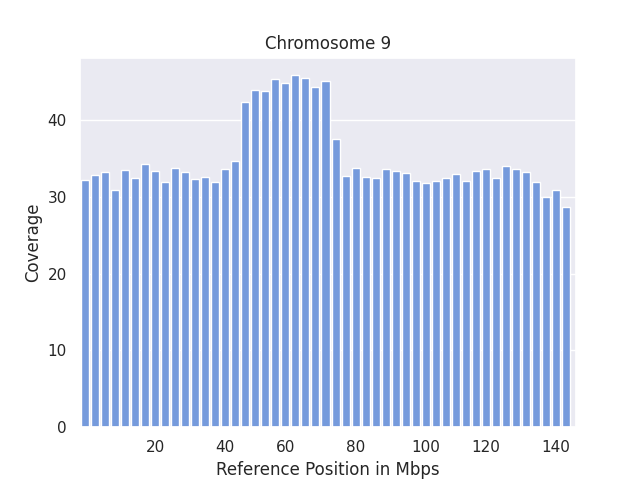} &
  \includegraphics[width=0.5\linewidth]{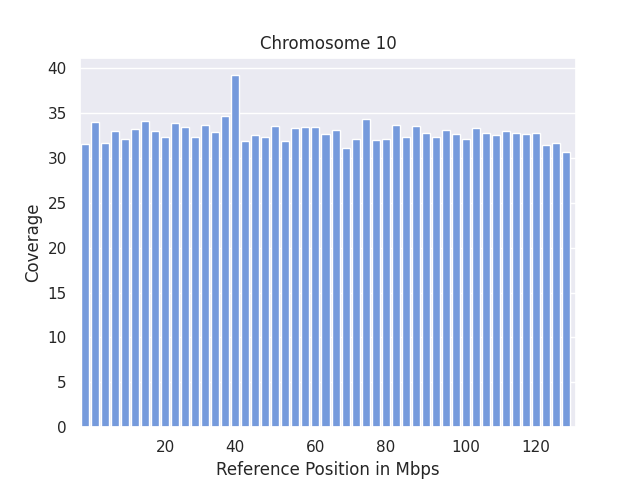} \\
  \includegraphics[width=0.5\linewidth]{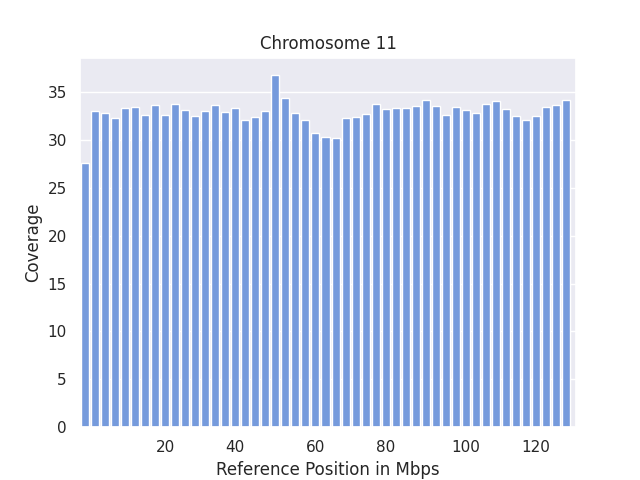} &
  \includegraphics[width=0.5\linewidth]{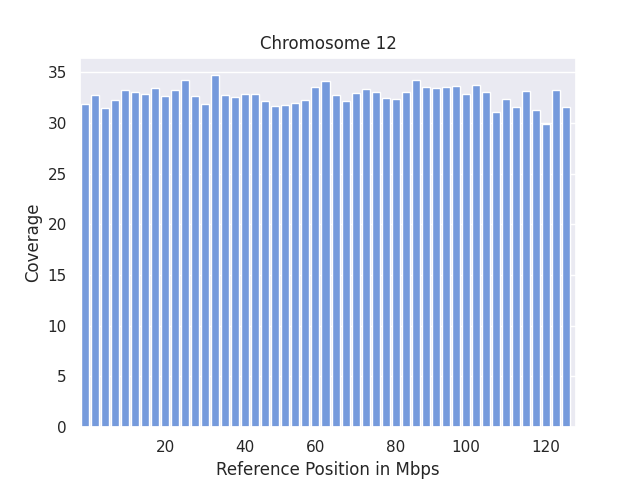} \\
    \includegraphics[width=0.5\linewidth]{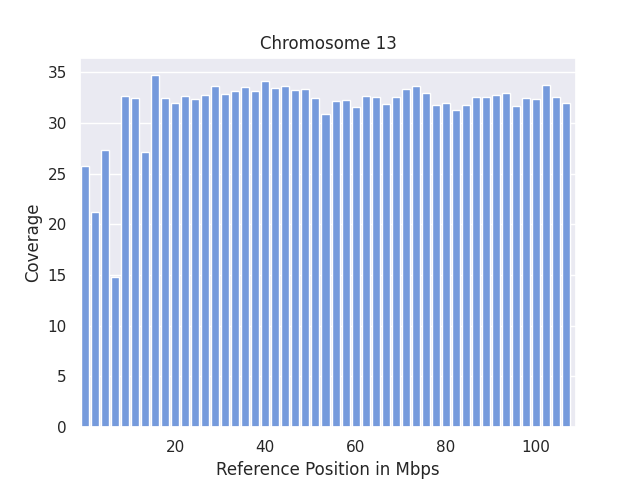} &
  \includegraphics[width=0.5\linewidth]{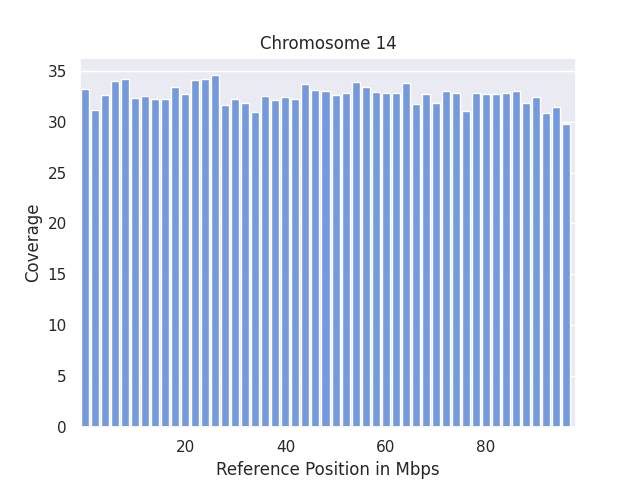} \\
  \includegraphics[width=0.5\linewidth]{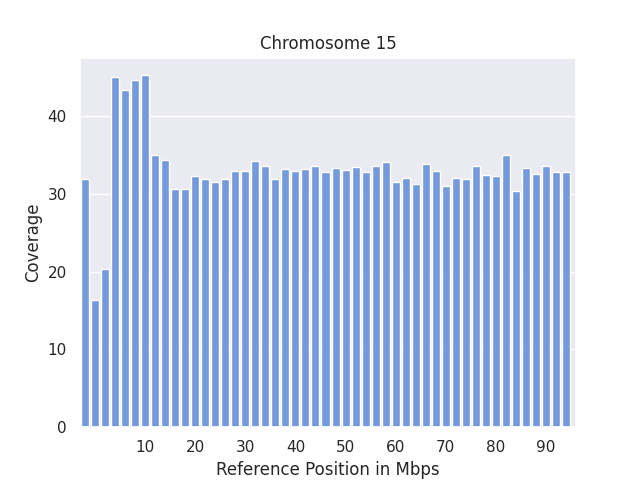} &
  \includegraphics[width=0.5\linewidth]{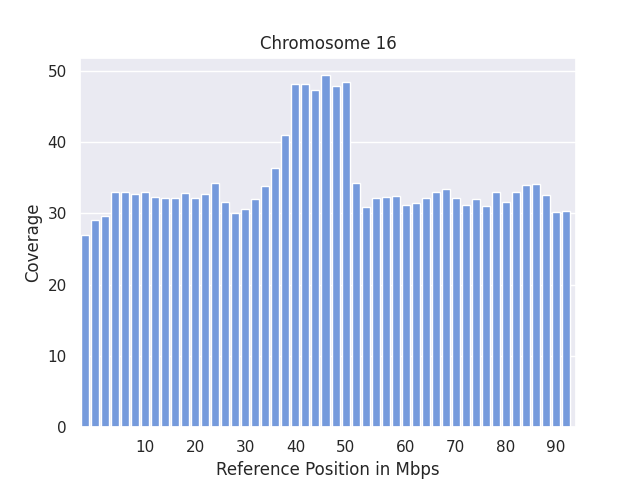} \\
\end{tabular}
\caption{Coverage histograms for real HiFi reads of chromosomes 9 to 16.}
\label{fig:cov2}
\end{figure}

\begin{figure}
\begin{tabular}{cc}

  \includegraphics[width=0.5\linewidth]{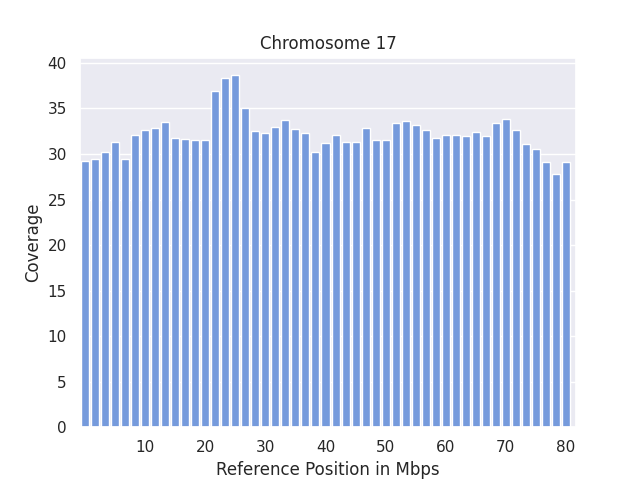} &
  \includegraphics[width=0.5\linewidth]{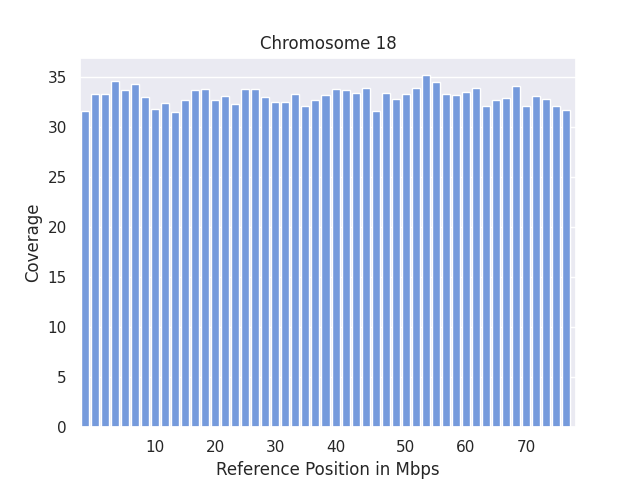} \\
  \includegraphics[width=0.5\linewidth]{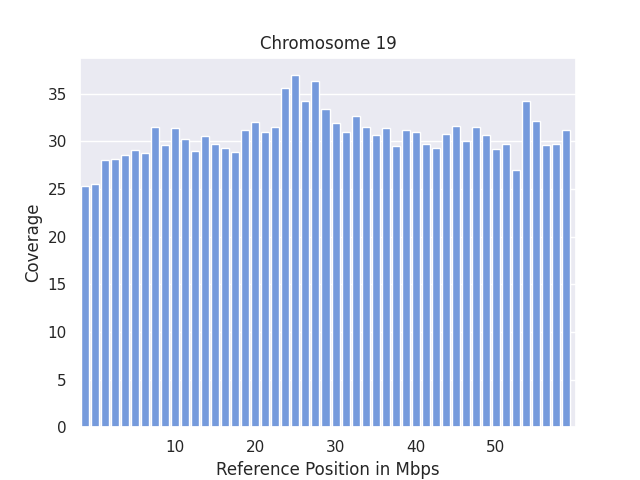} &
  \includegraphics[width=0.5\linewidth]{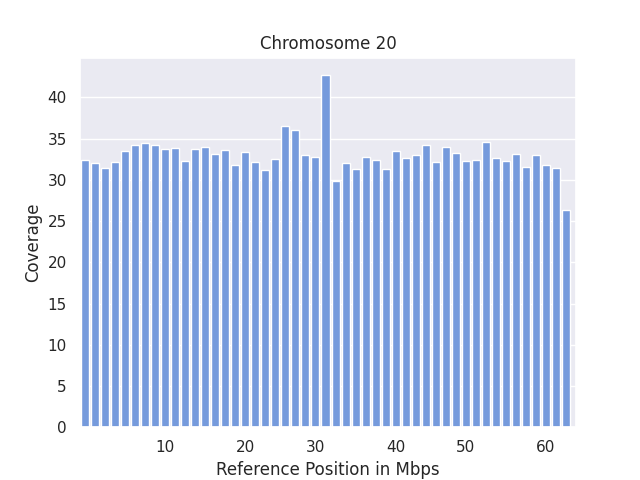} \\
    \includegraphics[width=0.5\linewidth]{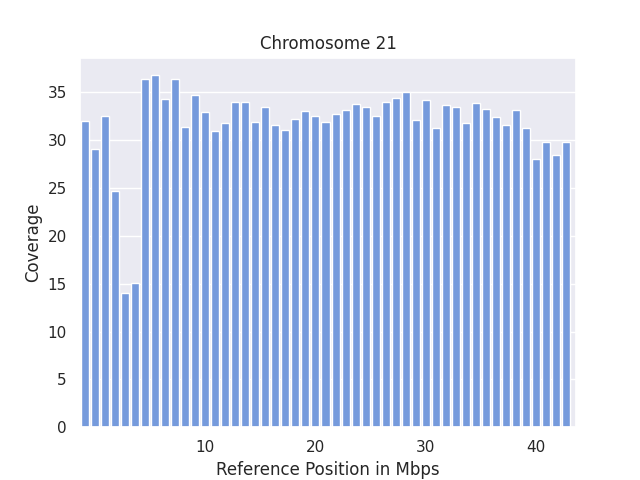} &
  \includegraphics[width=0.5\linewidth]{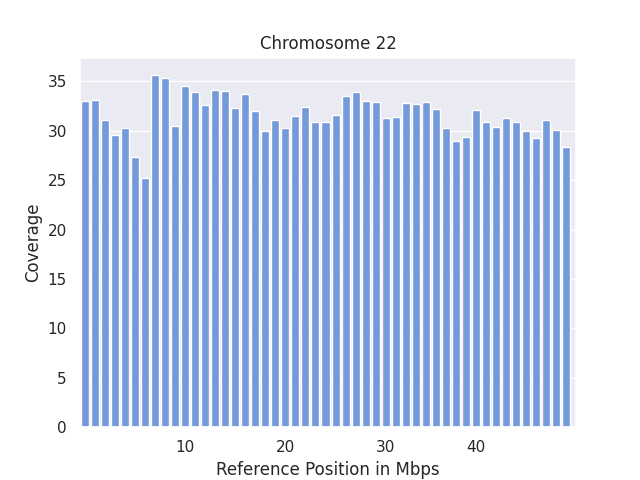} \\
  \includegraphics[width=0.5\linewidth]{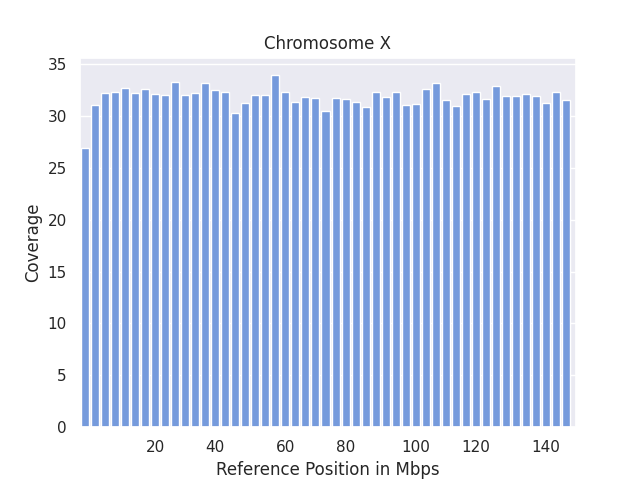} \\
\end{tabular}
\caption{Coverage histograms for real HiFi reads of chromosomes 17 to X.}
\label{fig:cov3}
\end{figure}

In the rest of this Section, we provide different statistics of the real PacBio HiFi reads we error-corrected and annotated by finding location on the chromosome for each read. These reads are automatically downloaded upon setting up the repository, as explained in Section \ref{supp:code} of these supplementary materials. We show histogram of read lengths of all the reads in the real dataset in Figure \ref{fig:readlength}, histogram of overlap lengths of all the overlaps in the real dataset in Figure \ref{fig:overlaplength}, as these plots do not vary significantly between different chromosomes. We also show coverage of each chromosome in real dataset---uneven coverage can indicate regions which are more difficult to assemble, such as repetitive regions inside centromeres and telomeres, and gives us insight into where the fragmentation is likely to happen. The coverage histograms are shown in Figures \ref{fig:cov1}, \ref{fig:cov2}, and \ref{fig:cov3}.

\end{document}